\documentclass[10pt,journal]{IEEEtran}

\usepackage{amsfonts}
\usepackage{amssymb}
\usepackage{amsthm}
\usepackage{amsmath,amsfonts,amssymb}
\usepackage[dvips]{graphicx}
\usepackage{verbatim}
\usepackage{setspace}
\usepackage{bm}
\usepackage{subfigure}
\usepackage{algorithmic} 
\usepackage[ruled,vlined]{algorithm2e}
\usepackage{cite}
\usepackage{enumerate}
\usepackage{indentfirst}
\usepackage{changepage}
\usepackage{pdfpages}
\usepackage{color}
\usepackage{lettrine}
\usepackage{url}

\IEEEoverridecommandlockouts

\begin{document}
	
	\title{Channel Estimation for Movable Antenna Aided Wideband Communication Systems}

		\author{
		Zhenyu Xiao,~\IEEEmembership{Senior Member,~IEEE,}
		Songqi Cao,~\IEEEmembership{Graduate Student Member,~IEEE,}
		Lipeng Zhu,~\IEEEmembership{Member,~IEEE,}
		Boyu Ning,~\IEEEmembership{Member,~IEEE,}
		Xiang-Gen Xia,~\IEEEmembership{Fellow,~IEEE,}
		and Rui Zhang,~\IEEEmembership{Fellow,~IEEE}

	\vspace{-0.5 cm}
		
		\thanks{ This work was supported in part by the National Natural Science Foundation of China (NSFC) under grant numbers 62171010 and U22A2007. (\emph{Corresponding author: Lipeng Zhu}) }
		\thanks{S.~Cao and Z.~Xiao are with the School of Electronic and Information Engineering, Beihang University, Beijing 100191, China. (e-mail: xiaozy@buaa.edu.cn, csq@buaa.edu.cn, liuyanming@buaa.edu.cn).}
		\thanks{L.~Zhu is with the Department of Electrical and Computer Engineering, National University of Singapore, Singapore 117583, Singapore. (e-mail: zhulp@nus.edu.sg).}
		\thanks{B.~Ning is with National Key Laboratory of Science and Technology on Communications, University of Electronic Science and Technology of China (UESTC), Chengdu 611731, China (e-mail: boydning@outlook.com).}
		\thanks{X.-G. Xia is with the Department of Electrical and Computer Engineering, University of Delaware, Newark, DE 19716, USA. (e-mail: xxia@ee.udel.edu).}
		\thanks{R. Zhang is with the School of Science and Engineering, Shenzhen Research Institute of Big Data, The Chinese University of
			Hong Kong, Shenzhen, Guangdong 518172, China (e-mail:rzhang@cuhk.edu.cn). He is also with the Department of Electrical
			and Computer Engineering, National University of Singapore, Singapore 117583 (e-mail: elezhang@nus.edu.sg).}
	}	
	
	\maketitle

	\begin{abstract}
		Movable antenna (MA) is an emerging technology that can significantly improve communication performance via the continuous adjustment of the antenna positions. To unleash the potential of MAs in wideband communication systems, acquiring accurate channel state information (CSI), i.e., the channel frequency responses (CFRs) between any position pair within the transmit (Tx) region and the receive (Rx) region across all subcarriers, is a crucial issue. In this paper, we study the channel estimation problem for wideband MA systems. To start with, we express the CFRs as a combination of the field-response vectors (FRVs), delay-response vector (DRV), and path-response tensor (PRT), which exhibit sparse characteristics and can be recovered by using a limited number of channel measurements at selected position pairs of Tx and Rx MAs over a few subcarriers. Specifically, we first formulate the recovery of the FRVs and DRV as a problem with multiple measurement vectors in compressed sensing (MMV-CS), which can be solved via a simultaneous orthogonal matching pursuit (SOMP) algorithm. Next, we estimate the PRT using the least-square (LS) method. Moreover, we also devise an alternating refinement approach to further improve the accuracy of the estimated FRVs, DRV, and PRT. This is achieved by minimizing the discrepancy between the received pilots and those constructed by the estimated CSI, which can be efficiently carried out by using the gradient descent algorithm. Finally, simulation results demonstrate that both the SOMP-based channel estimation method and alternating refinement method can reconstruct the complete wideband CSI with high accuracy, where the alternating refinement method performs better despite a higher complexity.
	\end{abstract}
	
	
	\section{Introduction}\label{section_1}

	With the development of wireless communications, the need for ultra-high reliability and capacity has grown radically. In this context, multiple-input multiple-output (MIMO) technology has been advanced to explore the degrees of freedom (DoFs) in the spatial domain more effectively \cite{6736761,8354789,4907446,9704317}. By exploiting the multiplexing and diversity gains in the spatial domain, the transmission rates and reliability can be significantly increased for MIMO systems \cite{10045774}. Nevertheless, since the antennas in conventional MIMO systems are discretely placed at fixed positions, the spatial DoFs cannot be fully utilized, which limits the performance of wireless communication systems. 
	
	To overcome this limitation, holographic MIMO (H-MIMO), also known as continuous-aperture MIMO or reconfigurable holographic surface (RHS), has been proposed to ultra-densely place tiny antennas in a specific region to exploit the maximum DoF in the spatial domain, which promotes the development of electromagnetic information theory (EIT) \cite{9724113,10163760,10417101,zhu2024mimo,zhu2023can}. However, since the elements are still discretely placed due to the hardware limitation, the DoFs in the spatial domain cannot be fully exploited. There still has a long way to resolve the technical challenges in H-MIMO and RHS technologies and apply them to existing communication systems. For example, the H-MIMO and RHS may undergo the coupling effect due to the sub-wavelength placement of the elements, which may bring extra interference to the desired signal \cite{9500830}. Another challenge lies in the limitation on processing broadband signals due to the lack of independent radio-frequency (RF) chain for each antenna element, leading to the degradation of the spatial multiplexing capability.
	
	Similar to the idea of exploring continuous spatial channel variation, movable antenna (MA) technology, also known as fluid-antenna system (FAS) \cite{wong2020fluid}, allows the antennas to move continuously in a specific region, which has recently been recognized as a promising technology to fully exploit the DoFs in the spatial domain. The continuous adjustment of the antennas' positions brings improvement in channel conditions, which thus significantly enhances the performance of communication systems. Specifically, compared to conventional FPA systems suffering from random fading, the MA-aided communication systems allow the MAs to search for positions with higher channel gains, thereby enhancing the received signal accordingly \cite{zhu2022modeling}. Similarly, by adjusting the MAs to the positions with high signal-to-interference-and-noise ratio (SINR), it is possible to simultaneously achieve interference mitigation and enhance the power of the desired signal \cite{zhu2023mov}. Moreover, by optimizing the positions of the MAs, the channel matrix can be effectively altered, which thus enhances the performance of MIMO spatial multiplexing \cite{ma2022mimo}. The practical implementation of MAs has also been preliminarily investigated \cite{zhu2024historical}, including motor-based \cite{zhu2023mov,8060521}, micro-electromechanical systems (MEMS)-based \cite{8041885}, fluid-based \cite{wong2020fluid}, and electronically driven architectures \cite{ning2024movable}, allowing the antennas to change their positions in different efficient ways under various communication scenarios.  
	
	Numerous studies have demonstrated the promising potential of MA-aided communication systems. The concept of MAs was introduced and applied to wireless communication systems in \cite{zhu2022modeling}, in which the field-response channel model was developed to characterize the continuous variation of wireless channels in confined regions. The performance improvement of MA systems was analyzed, showing that the MA-aided single-input single output (SISO) system can achieve a comparable signal-to-noise ratio (SNR) performance of a FPA-aided single-input multiple-output (SIMO) beamforming system. Moreover, it has been shown that with the increasing multipath component (MPC) number and region size, the performance of MA systems can be further enhanced. By optimizing the positions of the MAs at the transmitter (Tx), the received signal power was maximized for an MA-enhanced multiple-input single-output (MISO) system in \cite{10508218}, while the effective receive SNR was maximized for the MA-enabled communications with coordinated multi-point (CoMP) reception in \cite{10414081}. Beyond boosting received signal power, MA systems can also benefit from reduced interference. Specifically, the results in \cite{zhu2023mov} showed that the maximum SINR achieved by a single receive MA is close to the maximum SNR without jammers, indicating that the optimization of MAs can effectively suppress the interference with almost no loss of desired signal power. In \cite{10458417}, the inter-user interference as well as the transmit power was minimized via the optimization of the positions of the MAs at the Tx and the precoding matrix. Moreover, the applications of interference mitigation were studied in \cite{10447471,10416363} for secure communication and \cite{10579873} for cognitive radio, respectively. 
	
	In addition, spatial multiplexing gains can be improved by optimizing the positions of the MAs, which reshapes the channel matrix accordingly. The spatial multiplexing performance of MA systems was characterized in \cite{ma2022mimo}, where the channel capacity was maximized by jointly optimizing the positions of the MAs and the covariance matrix of the transmitted signal. In \cite{10437006}, the achievable rate was maximized by the joint optimization of the MAs and covariance matrix of the transmitted signal based on statistical channel state information (CSI). Moreover, the investigation of MAs has been extended to multi-user scenarios. With the optimization of the MAs' positions, the communication performance for multi-user scenarios can be significantly improved in terms of increasing the achievable rate of each user \cite{xiao2023multiuser}, reducing the total transmit power \cite{10354003}, and maximizing the multi-user capacity \cite{cheng2023sum,sun2023sum}. Moreover, with the aid of MA array, flexible multi-beam forming and null steering were studied in \cite{10382559} and \cite{10278220}, respectively. The application of MA-enabled flexible beamforming in satellite communications was investigated in \cite{zhu2024dynamic}. Given the above benefits, an overview of MA-aided narrowband communications was provided in \cite{zhu2023mov}. This overview encapsulates the advantages of MA systems in terms of signal power enhancement, interference reduction, beamforming capabilities, and spatial multiplexing. In addition to MA-aided narrowband communications, the implementation of MAs in the wideband communication systems was first studied in \cite{zhu2024performance}, in which a close-to-bound transmission rate for orthogonal frequency division multiplexing (OFDM) systems was achieved by jointly optimizing the positions of the MAs and power allocation of subcarriers.
	
	Although MA systems have great potentials in enhancing communication performance, such improvements rely on precise CSI. This underscores the significance of channel estimation to the efficacy of MA systems. In particular, given the MAs' ability to move flexibly within confined regions, the complete CSI, i.e., the channel responses between any position pair within the Tx region and the Rx region across all subcarriers, is required. Although there have been many studies on channel estimation such as \cite{7454701,7174558,8846224,8306126,7961152}, the methods therein are only designed for conventional FPA systems, and can only estimate the limited number of channel responses between discrete positions where the FPAs are located. In this regard, conventional channel estimation methods cannot be applied in MA-aided communication systems. Moreover, direct measurement of the channel responses incurs prohibitively high pilot overhead and is too time-consuming to be practically used. 
	
	In this regard, the channel estimation for MA-aided communication systems was studied \cite{zhang2023successive,10236898,10497534,ruo2024channel}. In particular, based on the assumption of strong spatial correlation, the authors in \cite{zhang2023successive} modeled the channel response as a Gaussian stochastic process. Then, by measuring the channel responses in the pre-defined positions, the complete CSI can be reconstructed via the maximum a posteriori (MAP) estimation. However, this method only considered the 1-dimensional (1D) movement of one antenna at the Rx, and extending to the 2-dimensional (2D) antenna movement at both the Tx and Rx sides may lead to extremely large training overhead. Thus, based on the field-response channel model, the successive transmitter-receiver compressed sensing (STRCS) method and joint channel estimation method were proposed in \cite{10236898} and \cite{10497534}, respectively. Specifically, both of these two methods exploited the effective path sparsity and formulated the channel estimation problem as a sparse signal recovery problem, in which the angles of departure (AoDs), angles of arrival (AoAs), and complex gains of each path are estimated via the compressed sensing method. The joint channel estimation method achieves more accurate CSI while the STRCS method has a lower computational complexity. In addition to the compressed sensing-based methods, the authors in \cite{ruo2024channel} proposed a tensor decomposition method to recover the AoDs, AoAs, and complex gains, which can achieve a significant improvement in channel estimation accuracy compared to the compressed sensing-based methods. However, the tensor decomposition method requires a regular structure of the MA measurement positions, which limits its generalization to other measurement setups and may result in a high measurement overhead. Despite the above works on channel estimation \cite{zhang2023successive,10236898,10497534,ruo2024channel}, they are tailored for narrowband MA systems, and may not be efficient in wideband MA systems. 
	
	Motivated by the above, in this paper, we study the channel estimation problem for wideband MA systems, in which the complete wideband CSI can be reconstructed by using a limited number of channel measurements at selected position pairs of Tx and Rx MAs over a few subcarriers. The main contributions of this paper are summarized as follows:
	\begin{itemize}
		\item An SOMP-based channel estimation framework for wideband MA systems is proposed. Specifically, we first decompose the channel frequency responses (CFRs) into the field-response vectors (FRVs), delay-response vector (DRV), and path-response tensor (PRT). Then, the recovery of the FRVs and DRV is formulated as a problem with multiple measurement vectors in compressed sensing (MMV-CS), which can be solved via a simultaneous orthogonal matching pursuit (SOMP) algorithm. Next, the PRT is estimated via the least-square (LS) method. Based on the estimated FRVs, DRV, and PRT, the complete CSI can be reconstructed. 
		\item An alternating refinement method is developed to further improve the accuracy of the estimated FRVs, DRV, as well as PRT. To achieve this end, we formulate an optimization problem aimed at minimizing the discrepancy between the received pilot signals and those constructed by the estimated CSI, which can be solved via the gradient descent algorithm. Accordingly, the accuracy of channel estimation can be further improved. 
		\item Extensive simulations are carried out to verify the effectiveness of the proposed methods. The results indicate that both the SOMP-based channel estimation method and the alternating refinement method are capable of reconstructing the complete CSI with high accuracy, where the alternating refinement method performs better although it entails a higher complexity. 
	\end{itemize} 

	The rest of this paper is organized as follows. Section \ref{section_2} introduces the system model and the wideband field-response channel model for the considered MA system. In Section \ref{section_3}, we develop the SOMP-based channel estimation method, while the alternating refinement method is introduced in Section \ref{section_4}. Simulation results are presented in Section \ref{section_5} and this paper is finally concluded in Section \ref{section_6}. 
	
	\emph{Notation}: $a, {\bf a}, {\bf A}$, and $\mathcal{A}$ denote a scalar, a vector, a matrix, and a set, respectively. $\left[\bf a\right]_{i}$ represents the $i$-th element of vector $\bf a$. $\left[{\bf A}\right]_{i,:}$, $\left[{\bf A}\right]_{:,j}$, and $\left[{\bf A}\right]_{i,j}$ represent the $i$-th row, the $j$-th column, and the $i$-th row and $j$-th column element of matrix ${\bf A}$, respectively. $\left(\cdot\right)^{*}$, $\left(\cdot\right)^{\rm T}$ and $\left(\cdot\right)^{\rm H}$ denote conjugate, transpose and conjugate transpose operations, respectively. We use $\times_1$, $\times_2$, and $\times_3$ to represent the $1$-mode, 2-mode, and 3-mode products of a tensor, respectively. Denote ${\rm vec}\left(\cdot\right)$, $\lceil \cdot \rceil$, and $\bmod\left(\cdot\right)$ as the vectorization, ceiling, and modulo  operations, respectively. In addition, $\otimes$ denotes the Kronecker product. $\mathcal{CN}\left(0,\sigma^2\right)$ represents the circularly symmetric complex Gaussian (CSCG) distribution with mean zero and variance $\sigma^2$. $\mathbb{C}$ and $\mathbb{Z}$ denote the sets of complex numbers and integers, respectively. $\Vert \cdot \Vert_2$, and $\Vert \cdot \Vert_{\rm F}$ denote the $l_2$-norm and Frobenius norm, respectively. We use ${\bf 1}_{N}$ to represent the all $1$ vector of dimension $N$. 
	
	\section{System and Channel Model}\label{section_2}

	\subsection{System Model}

		\begin{figure*}[htbp]
		\centering
		\includegraphics[width=15 cm]{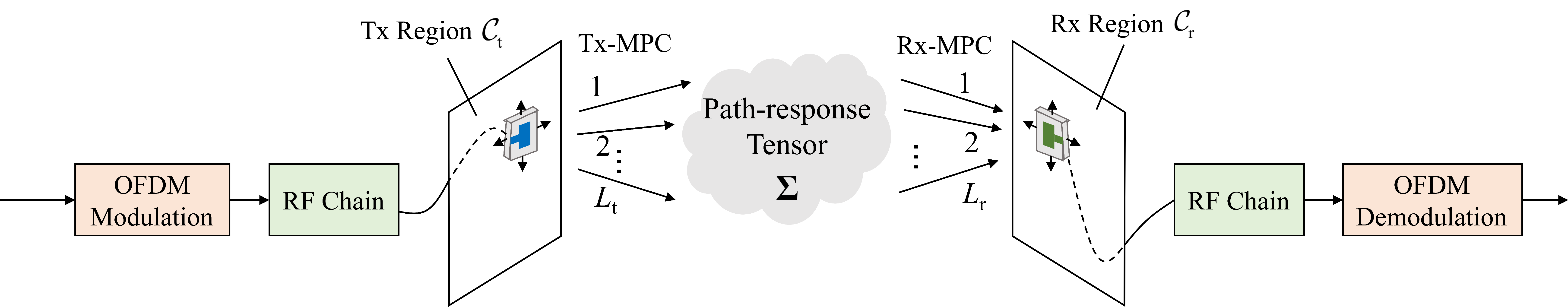}
		\caption{Illustration of the considered wideband MA communication system.}
		\label{fig:ma_system}
	\end{figure*}
	
	In this paper, an MA-OFDM communication system with $K$ subcarriers is considered. Denoting $T_{\rm s}$ as the sampling period, the system bandwidth is thus given by $1/T_{\rm s}$. As shown in Fig. \ref{fig:ma_system}, a single transmit MA (Tx-MA) and a single receive MA (Rx-MA) are deployed at Tx and Rx for transmitting and receiving signals, respectively. The Tx-MA and Rx-MA can flexibly change their positions in the Tx region $\mathcal{C}_{\rm t}$ and Rx region $\mathcal{C}_{\rm r}$, respectively. We assume that both Tx and Rx regions are square regions of size $S\lambda\times S\lambda$, in which $\lambda$ represents the wavelength corresponding to the carrier frequency of the system. The local Cartesian coordinate systems, $x_{\rm t}$-$O_{\rm t}$-$y_{\rm t}$ and $x_{\rm r}$-$O_{\rm r}$-$y_{\rm r}$, are established to represent the positions of the Tx-MA ${\bf t} = \left[x_{\rm t}, y_{\rm t}\right]^{\rm T}\in\mathcal{C}_{\rm t}$ and Rx-MA ${\bf r} = \left[x_{\rm r}, y_{\rm r}\right]^{\rm T}\in\mathcal{C}_{\rm r}$, respectively. 
	
	For MA systems, the received signal at the $k$-th subcarrier of each OFDM symbol can be represented as
	\begin{equation}\label{rec_signal}
		r\left({\bf t},{\bf r},k\right) = \sqrt{p_{\rm t}}h\left({\bf t,r}, k\right)s\left(k\right)+n\left(k\right),
	\end{equation}
	where $s\left(k\right)$ is the transmitted signal, $h\left({\bf t,r}, k\right)$ denotes the CFR between Tx-MA's position ${\bf t}$ and Rx-MA's position ${\bf r}$. Scalars $p_{\rm t}$ and $n\left(k\right)\in\mathcal{CN}\left(0,\sigma^2\right)$ represent the transmit power and the complex additive Gaussian noise with power $\sigma^2$ at the $k$-th subcarrier, respectively. 
	
	\subsection{Field-response Channel Model}
	
	For MA systems, CFR $h\left({\bf t, r}, k\right)$ is influenced by the movements of the Tx-MA and Rx-MA. In other words, the CFR is a function that varies continuously with the positions of the MAs. In the following, we focus on the characterization of the CFR for MA systems, i.e., $h\left({\bf t}, {\bf r}, k\right)$, by extending the field-response channel model, which is generally utilized in narrowband MA systems, to wideband MA systems \cite{zhu2022modeling}. Specifically, for a wireless channel with $L_{\rm t}$ transmit MPCs (Tx-MPCs), $L_{\rm r}$ receive MPCs (Rx-MPCs), and $L_{\rm d}$ delays (normalized by the sampling period $T_{\rm s}$), there are at most $L_{\rm t}\times L_{\rm r} \times L_{\rm d}$ MPCs, which can be represented by a path-response tensor (PRT) ${\bf \Sigma}\in\mathbb{C}^{L_{\rm r}\times L_{\rm t}\times L_{\rm d}}$. The row-index, column-index, and tube-index of ${\bf \Sigma}$ correspond to a unique pair of elevation and azimuth AoAs, a unique pair of elevation and azimuth AoDs, and a unique delay, respectively. Each element in ${\bf \Sigma}$ denotes the complex gain of a specific propagation path. Notably, a zero element in ${\bf \Sigma}$ indicates that there is no path with corresponding AoDs, AoAs, and delay. 
	
	Specifically, with the definition of PRT, the CFR at the $k$-th subcarrier with both the Tx-MA and Rx-MA fixed at the reference positions, i.e., ${\bf t}_{0} = \left[0,0\right]^{\rm T}$ and ${\bf r}_{0} = \left[0,0\right]^{\rm T}$, can be written as
	\begin{equation}\label{cfr_reference}
		\begin{aligned}
			h\left({\bf t}_0, {\bf r}_0,k\right) = &{\bf 1}_{L_{\rm r}}^{\rm H}\left({\bf \Sigma}\times_3 {\bf d}^{\rm T}\left(k\right)\right){\bf 1}_{L_{\rm t}}\\
			\overset{\left({\rm a}\right)}{=}&\sum_{l_{\rm d}=1}^{L_{\rm d}}\left(\sum_{l_{\rm t} = 1}^{L_{\rm t}}\sum_{l_{\rm r} = 1}^{L_{\rm r}}\left[{\bf \Sigma}\right]_{l_{\rm r}, l_{\rm t}, l_{\rm d}}\right) e^{-j2\pi\frac{k\tau^{l_{\rm d}}}{KT_{\rm s}}},
		\end{aligned}
	\end{equation}
	in which $\left[{\bf \Sigma}\right]_{l_{\rm r},l_{\rm t},l_{\rm d}}$ is the element in the $l_{\rm r}$-th row, $l_{\rm t}$-th column, and $l_{\rm d}$-th tube of PRT ${\bf \Sigma}$. Vector ${\bf d}\left(k\right)\in\mathbb{C}^{L_{\rm d}\times 1}$ is defined as the DRV, representing the phase shift at the $k$-th subcarrier w.r.t. the $0$-th subcarrier due to the delays. The DRV ${\bf d}\left(k\right), k = 0,1,\cdots,K-1,$ is given by
	\begin{equation}
		{\bf d}\left(k\right) = \left[e^{-j2\pi \frac{k\tau^{1}}{KT_{\rm s}}},\cdots,e^{-j2\pi \frac{k\tau^{L_{\rm d}}}{KT_{\rm s}}}\right]^{\rm T},
	\end{equation}
	in which $\left\{\tau^{l_{\rm d}}\right\}_{l_{\rm d} = 1}^{L_{\rm d}}$ are the delays. Notably, the channel gain of a path may decrease greatly by the long propagation distance with a large delay. In other words, the delays of the dominant propagation paths are no longer than a maximum delay $\tau_{\rm max}$, i.e., $\tau^{l_{\rm d}}<\tau_{\rm max}, l_{\rm d} = 1,\cdots, L_{\rm d}$ \cite{3gpp38901}. Moreover, step (a) in \eqref{cfr_reference} indicates that when the Tx-MA and Rx-MA are fixed at the reference positions, all the propagation paths can be regarded as $L_{\rm d}$ effective paths with complex gains $\left\{\sum_{l_{\rm t} = 1}^{L_{\rm t}}\sum_{l_{\rm r} = 1}^{L_{\rm r}}\left[{\bf \Sigma}\right]_{l_{\rm r}, l_{\rm t}, l_{\rm d}}\right\}_{l_{\rm d} = 1}^{L_{\rm d}}$.
	
\begin{figure}[t]
		\centering
		\includegraphics[width= 6 cm]{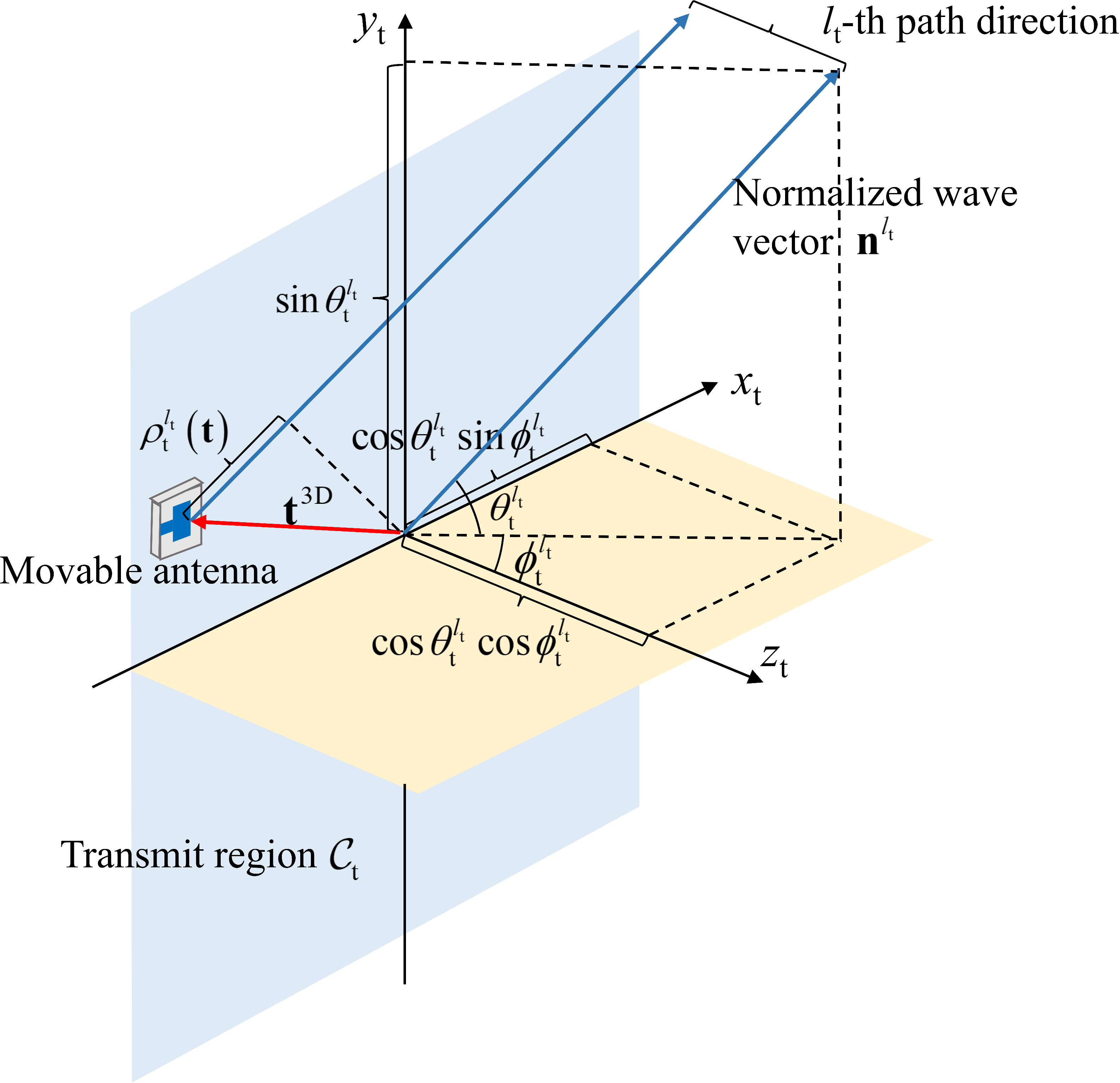}
		\caption{The coordinate system of the Tx region, in which the $l_{\rm t}$-th Tx-MPC is represented to illustrate the impact of the movement of the Tx-MA.}
		\label{fig:ma_system2}
	\end{figure}

	Next, we consider the impact of MAs' movements on the CFR. Denote $\theta_{\rm t}^{l_{\rm t}}\in\left[-\pi/2, \pi/2\right]$ and $\phi_{\rm t}^{l_{\rm t}}\in\left[-\pi/2,\pi/2\right]$ as the elevation and azimuth AoDs of the $l_{\rm t}$-th Tx-MPC, respectively. Then, the phase difference corresponding to the $l_{\rm t}$-th Tx-MPC between the position of the Tx-MA ${\bf t}$ and the reference position ${\bf t}_0$ is $2\pi\rho_{\rm t}^{l_{\rm t}}\left({\bf t}\right)/\lambda$, where $\rho_{\rm t}^{l_{\rm t}}\left({\bf t}\right)=x_{\rm t}\cos\theta_{\rm t}^{l_{\rm t}}\sin\phi_{\rm t}^{l_{\rm t}}+y_{\rm t}\sin\theta_{\rm t}^{l_{\rm t}}$, as shown in Fig. \ref{fig:ma_system2}. Similarly, the elevation and azimuth AoAs of the $l_{\rm r}$-th Rx-MPC is denoted by $\theta_{\rm r}^{l_{\rm r}}\in\left[-\pi/2, \pi/2\right]$ and $\phi_{\rm r}^{l_{\rm r}}\in\left[-\pi/2,\pi/2\right]$, respectively, and the phase difference for the $l_{\rm r}$-th Rx-MPC between position ${\bf r}$ and reference position at the Rx is $2\pi\rho_{\rm r}^{l_{\rm r}}\left({\bf r}\right)/\lambda$, where $\rho_{\rm r}^{l_{\rm r}} = x_{\rm r}\cos\theta_{\rm r}^{l_{\rm r}}\sin\phi_{\rm r}^{l_{\rm r}}+y_{\rm r}\sin\theta_{\rm r}^{l_{\rm r}}$.Then, the CFR at the $k$-th subcarrier with Tx-MA's position ${\bf t}$ and Rx-MA's position ${\bf r}$ is given by
	\begin{equation}\label{CFR}
		\begin{aligned}
			h\left({\bf t,r},k\right) &= {\bf f}\left({\bf r}\right)^{\rm H}\left({\bf \Sigma}\times_3 {\bf d}\left(k\right)\right){\bf g}\left({\bf t}\right)\\
			&= \left[\left({\bf \Sigma}\times_1 {\bf f\left(r\right)}^{\rm H}\right)\times_2 {\bf g\left(t\right)}^{\rm T}\right]\times_3 {\bf d}\left(k\right)^{\rm T},
		\end{aligned}
	\end{equation}
	in which ${\bf g}\left({\bf t}\right)\in\mathbb{C}^{L_{\rm t}\times 1}$ and ${\bf f}\left({\bf r}\right)\in\mathbb{C}^{L_{\rm r}\times 1}$ are defined as the transmit FRV (Tx-FRV) and receive FRV (Rx-FRV), respectively, i.e.,
	\begin{equation}
		\left\{
		\begin{aligned}
			{\bf g\left(t\right)}=\left[e^{j\frac{2\pi}{\lambda}\rho_{\rm t}^{1}\left({\bf t}\right)},\cdots,e^{j\frac{2\pi}{\lambda}\rho_{\rm t}^{L_{\rm t}}\left({\bf t}\right)}\right]^{\rm T},\\
			{\bf f\left(r\right)}=\left[e^{j\frac{2\pi}{\lambda}\rho_{\rm r}^{1}\left({\bf r}\right)},\cdots,e^{j\frac{2\pi}{\lambda}\rho_{\rm r}^{L_{\rm r}}\left({\bf r}\right)}\right]^{\rm T}.
		\end{aligned}
		\right.
	\end{equation}
	It can be observed in \eqref{CFR} that the CFR can be decomposed into the Tx-FRV, Rx-FRV, DRV, and PRT, containing the AoDs, AoAs, delays, and complex gains, namely wideband MPC information, which exhibits sparse characteristic. Specifically, compared to the CFRs, the propagation paths and the corresponding wideband MPC information are finite, thus recovering the wideband MPC information allows us to reconstruct the complete wideband CSI.
	
	Substituting \eqref{CFR} into \eqref{rec_signal}, the received signal at the $k$-th subcarrier can be represented as
	\begin{equation}
		\begin{aligned}
			&r\left({\bf t,r}, k\right)=\\
			&\sqrt{p_{\rm t}}\left\{\left[\left({\bf \Sigma}\times_1 {\bf f\left(r\right)}^{\rm H}\right)\times_2 {\bf g\left(t\right)}^{\rm T}\right]\times_3 {\bf d}\left(k\right)^{\rm T}\right\}s\left(k\right)+n\left(k\right).
		\end{aligned}
	\end{equation}
	
	The objective of channel estimation for MA systems is to recover the complete wideband CSI, i.e., the CFR $h\left({\bf t,r},k\right)$ between any position pair within the Tx region and the Rx region across all the $K$ subcarriers. Directly traversing the whole Tx and Rx regions may result in a high pilot overhead and time consumption, which is impractical in reality. Fortunately, in \eqref{CFR}, the CFR is decomposed into FRVs, DRV, and PRT, in which the corresponding wideband MPC information exhibits sparse characteristic and can be estimated via a limited number of pilots. Motivated by this, in this paper, we consider the channel estimation method by estimating the wideband MPC information.

	\section{SOMP-based Channel Estimation}\label{section_3}
	
	In this section, we develop the SOMP-based channel estimation method using the sparse characteristic of the propagation paths and the corresponding MPC information. Specifically, based on the proposed pilot scheme, the estimation of the FRVs and DRV is formulated as a problem of MMV-CS and solved via the SOMP algorithm. Next, the PRT is estimated via the LS method. Then, based on the field-response channel model in \eqref{CFR}, the complete wideband CSI can be reconstructed. 
	
\begin{figure}[t]
		\centering
		
		\subfigure[]{
			\begin{minipage}[t]{1\linewidth}
				\centering
				\includegraphics[width= 8.5 cm]{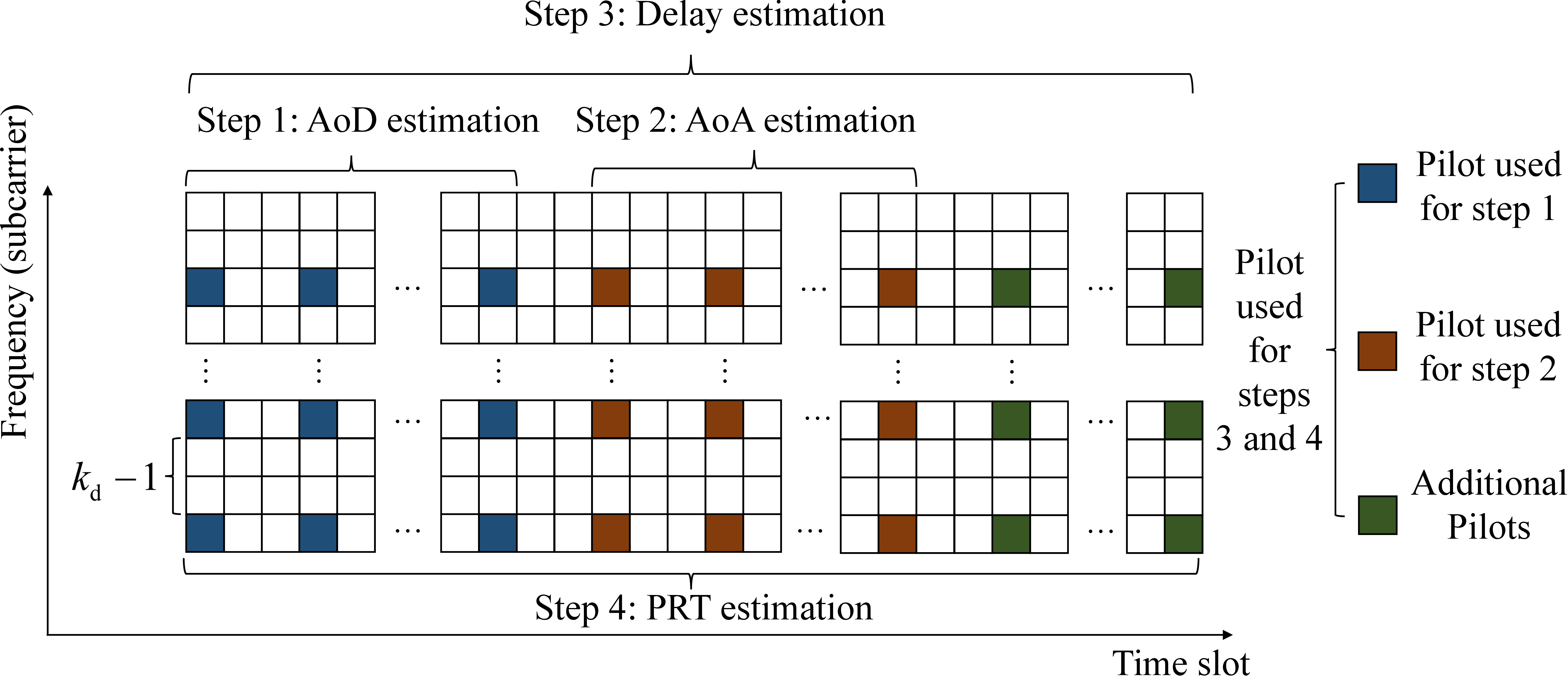}
				\label{fig:ma_pilot}
			\end{minipage}%
		}%
		\\
		\subfigure[]{
			\begin{minipage}[t]{1\linewidth}
				\centering
				\includegraphics[width= 7.5 cm]{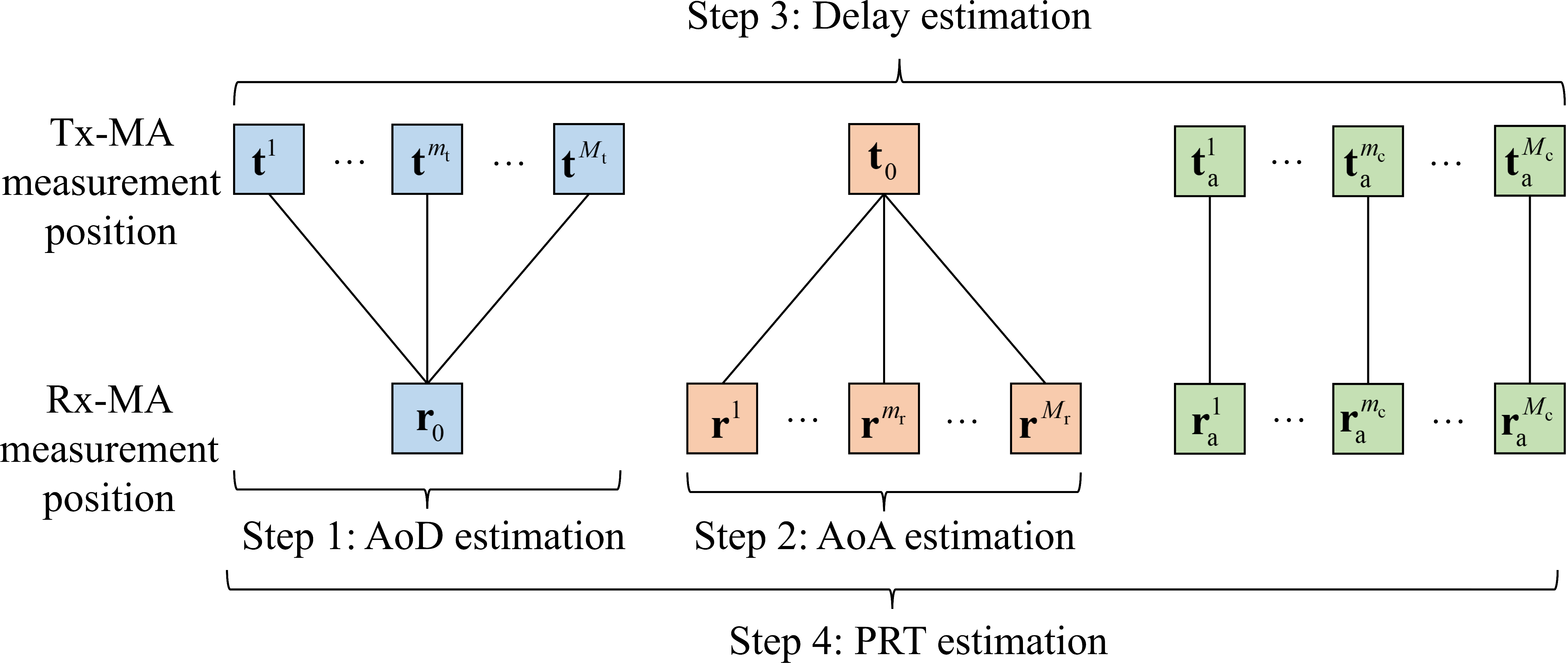}
				\label{fig:ma_position}
			\end{minipage}%
		}
		
		\centering
		\caption{Illustration of the (a) proposed pilot scheme and (b) MA movement scheme for channel estimation.}
	\end{figure}

	For channel estimation, CFR measurements are performed by transmitting OFDM symbols with pilots, namely pilot symbols. For one pilot symbol, $M_{\rm d}$ pilots are uniformly distributed across all the $K$ subcarriers, and adjacent pilots are separated by $k_{\rm d}-1$ subcarriers, as shown in Fig. \ref{fig:ma_pilot}. In other words, the pilots are inserted at the $k_{\rm d}\left(m_{\rm d}-1\right)$-th subcarrier, $m_{\rm d} = 1,\cdots, M_{\rm d}$, and $k_{\rm d}M_{\rm d} = K$ is satisfied. For notation simplicity, we define $k_{m_{\rm d}}=k_{\rm d}\left(m_{\rm d}-1\right), m_{\rm d} = 1,\cdots, M_{\rm d}$.
	
	For MA systems, the positions of MAs can be flexibly adjusted when transmitting different pilot symbols. In our proposed channel estimation framework, the movements of MAs are implemented in three steps, as shown in Fig. \ref{fig:ma_position}. In step 1, the Tx-MA moves over different positions while the Rx-MA is fixed at reference position ${\bf r}_0 = \left[0,0\right]^{\rm T}$ to perform AoD estimation. In step 2, the Rx-MA moves over different positions while the Tx-MA is fixed at reference position ${\bf t}_0 = \left[0,0\right]^{\rm T}$ to perform AoA estimation. In step 3, the Tx-MA and Rx-MA move simultaneously for additional channel measurements, which guarantees the recovery of the PRT \cite{10236898}. Based on the channel measurements in all the three steps, the delay estimation is performed based on the received pilots of all the three steps. Finally, the PRT is estimated via LS method based on the estimated AoDs, AoAs, and delays.
	
	\subsection{AoD and AoA Estimation}\label{angle_estimation}

	For AoD estimation, the Tx-MA moves to different positions and transmits pilot symbols to the Rx-MA for channel measurement, which is fixed at reference position ${\bf r}_0$. Denote ${\bf t}^{m_{\rm t}} = \left[x_{\rm t}^{m_{\rm t}}, y_{\rm t}^{m_{\rm t}}\right]^{\rm T}, m_{\rm t} = 1,\cdots,M_{\rm t}$ as the measurement positions of the Tx-MA, in which $M_{\rm t}$ denotes the total number of the measurement positions in step 1. For each position ${\bf t}^{m_{\rm t}}$, we assume that one pilot symbol is transmitted by the Tx-MA. Then, the received signal at the $k_{m_{\rm d}}$-th, $1\leq m_{\rm d} \leq M_{\rm d}$, subcarrier can be represented as
	\begin{equation}
		\begin{aligned}
			&v\left({\bf t}^{m_{\rm t}},{\bf r}_0,k_{m_{\rm d}}\right) \\= 
			&\sqrt{p_{\rm t}}\left\{\left[\left({\bf \Sigma}\times_1{\bf f}\left({\bf r}_0\right)^{\rm H}\right)\times_2 {\bf g}\left({\bf t}^{m_{\rm t}}\right)^{\rm T}\right]\times_3 {\bf d}\left(k_{m_{\rm d}}\right)^{\rm T}\right\}s\left(k_{m_{\rm d}}\right)\\&+
			n^{m_{\rm t}}\left(k_{m_{\rm d}}\right),
		\end{aligned}
	\end{equation}
	in which $n^{m_{\rm t}}\left(k_{m_{\rm d}}\right)$ represents the noise in the received pilot at the $k_{m_{\rm d}}$-th subcarrier for the $m_{\rm t}$-th Tx-MA measurement position. For notation simplicity, we denote ${\bf X}_{\rm td} \triangleq {\bf \Sigma}\times_1{\bf f}\left({\bf r}_0\right)^{\rm H}\in\mathbb{C}^{L_{\rm t}\times L_{\rm d}} $. Without loss of generality, we assume the pilot $s\left(k_{m_{\rm d}}\right)=1$. Then, the received pilot at the $k_{m_{\rm d}}$-th subcarrier can be represented as
	\begin{equation}
		\begin{aligned}
			v\left({\bf t}^{m_{\rm t}}, {\bf r}_0,k_{m_{\rm d}}\right) =
			\sqrt{p_{\rm t}}{\bf g}\left({\bf t}^{m_{\rm t}}\right)^{\rm T}{\bf X}_{\rm td}{\bf d}\left(k_{m_{\rm d}}\right)+n^{m_{\rm t}}\left(k_{m_{\rm d}}\right).
		\end{aligned}
	\end{equation}
	Collecting all the $M_{\rm d}$ received pilots and stacking them into a vector, i.e., ${\bf v}_{\rm t}^{m_{\rm t}} = \left[v\left({\bf t}^{m_{\rm t}},{\bf r}_0,0\right),\cdots,v\left({\bf t}^{m_{\rm t}},{\bf r}_0,k_{m_{\rm d}}\right)\right]\in\mathbb{C}^{1\times M_{\rm d}}$, we have
	\begin{equation}
		{\bf v}_{\rm t}^{m_{\rm t}} = \sqrt{p_{\rm t}}{\bf g}\left({\bf t}^{m_{\rm t}}\right)^{\rm T}{\bf X}_{\rm td}{\bf D}+{\bf n}_{\rm t}^{m_{\rm t}},
	\end{equation}
	in which ${\bf D} = \left[{\bf d}\left(0\right),\cdots,{\bf d}\left(k_{m_{\rm d}}\right)\right]\in\mathbb{C}^{L_{\rm d}\times M_{\rm d}}$ and ${\bf n}_{\rm t}^{m_{\rm t}}=\left[n^{m_{\rm t}}\left(0\right),\cdots,n^{m_{\rm t}}\left(k_{m_{\rm d}}\right)\right]\in\mathbb{C}^{1\times M_{\rm d}}$. Then, collecting all the received pilots at all the $M_{\rm t}$ Tx-MA measurement positions, we have
	\begin{equation}\label{received_pilots_tx}
		{\bf V}_{\rm t} = \sqrt{p_{\rm t}}{\bf G}^{\rm T}{\bf X}_{\rm td}{\bf D}+{\bf N}_{\rm t},
	\end{equation}
	in which ${\bf G} = \left[{\bf g}\left({\bf t}^{1}\right),\cdots,{\bf g}\left({\bf t}^{M_{\rm t}}\right)\right]\in\mathbb{C}^{L_{\rm t}\times M_{\rm t}}$ and ${\bf N}_{\rm t} = \left[\left({\bf n}_{\rm t}^{1}\right)^{\rm T},\cdots,\left({\bf n}_{\rm t}^{M_{\rm t}}\right)^{\rm T}\right]^{\rm T}\in\mathbb{C}^{M_{\rm t}\times M_{\rm d}}$. 
	
	Since virtual AoDs $\left\{\vartheta_{\rm t}^{l_{\rm t}}, \varphi_{\rm t}^{l_{\rm t}}\right\}_{l_{\rm t} = 1}^{L_{\rm t}}$ can be any values within $-1$ to $1$, which contain infinite numbers of possible values, the virtual AoDs cannot be perfectly recovered via a limited number of pilots. In this regard, we discretize the AoDs. Specifically, define sets $\varTheta_{\rm t}$ and $\varPhi_{\rm t}$ as
	\begin{equation}
		\begin{small}
			\left\{
		\begin{aligned}
			\varPhi_{\rm t} = \left\{\tilde{\varphi}_{\rm t}^{g_{tx}}=-1+\frac{2g_{tx}-1}{G}\left|1\leq g_{tx}\leq G\right.\right\},\\
			\varTheta_{\rm t} = \left\{\tilde{\vartheta}_{\rm t}^{g_{ty}}=-1+\frac{2g_{ty}-1}{G}\left|1\leq g_{ty}\leq G\right.\right\},
		\end{aligned}\right.
		\end{small}
	\end{equation}
	to approximate all the virtual AoDs. The discretization of the AoDs leads to a discretization error ${\bf E}_{\rm t}\in\mathbb{C}^{M_{\rm t}\times M_{\rm d}}$ due to the mismatch between the discretized AoDs in sets $\varPhi_{\rm t}$ and $\varTheta_{\rm t}$ and continuous angles $\left\{\varphi_{\rm t}^{l_{\rm t}}, \vartheta_{\rm t}^{l_{\rm t}}\right\}_{l_{\rm t}}^{L_{\rm t}}$. Consequently, \eqref{received_pilots_tx} becomes
	\begin{equation}
		{\bf V}_{\rm t} = \sqrt{p_{\rm t}}\tilde{{\bf G}}^{\rm T}\tilde{{\bf X}}_{\rm td}{\bf D}+{\bf E}_{\rm t}+{\bf N}_{\rm t}\triangleq\sqrt{p_{\rm t}}\tilde{{\bf G}}^{\rm T}\tilde{{\bf U}}_{\rm t}+{\bf E}_{\rm t}+{\bf N}_{\rm t},
	\end{equation}
	with $\tilde{{\bf G}} = \left[\tilde{\bf g}\left({\bf t}^1\right),\cdots,\tilde{\bf g}\left({\bf t}^{M_{\rm t}}\right)\right]\in\mathbb{C}^{G^2\times M_{\rm t}}$ and $\tilde{{\bf X}}_{\rm td}\in\mathbb{C}^{G^2\times M_{\rm d}}$. Vector $\tilde{\bf g}\left({\bf t}^{m_{\rm t}}\right)\in\mathbb{C}^{G^2\times 1}, m_{\rm t} = 1,\cdots,M_{\rm t},$ is defined as the discretized Tx-FRV, i.e., 
	\begin{equation}\label{discrete_frv_tx}
		\begin{small}
			\begin{aligned}
		\tilde{\bf g}\left({\bf t}^{m_{\rm t}}\right) =& \left[e^{j\frac{2\pi}{\lambda}y_{\rm t}^{m_{\rm t}}\tilde{\vartheta}_{\rm t}^{1}},\cdots,e^{j\frac{2\pi}{\lambda}y_{\rm t}^{m_{\rm t}}\tilde{\vartheta}_{\rm t}^{G_{ty}}}\right]^{\rm T}\\
		&\otimes\left[e^{j\frac{2\pi}{\lambda}x_{\rm t}^{m_{\rm t}}\tilde{\varphi}_{\rm t}^{1}},\cdots,e^{j\frac{2\pi}{\lambda}x_{\rm t}^{m_{\rm t}}\tilde{\varphi}_{\rm t}^{G_{tx}}}\right]^{\rm T}.
		\end{aligned}
		\end{small}
	\end{equation}
	Moreover, for notation simplicity, define $\tilde{{\bf U}}_{\rm t} \triangleq \tilde{{\bf X}}_{\rm td}{\bf D}\in\mathbb{C}^{G^2\times M_{\rm d}}$. It is worth noting that each row index of $\tilde{{\bf U}}_{\rm t}$ corresponds to one pair of discretized virtual AoDs defined in sets $\varPhi_{\rm t}$ and $\varTheta_{\rm t}$. Specifically, the $g_{\rm t}$-th row in $\tilde{{\bf U}}_{\rm t}$ corresponds to virtual AoDs $\varphi_{\rm t}^{g_{tx}}$ and $\vartheta_{\rm t}^{g_{ty}}$, which satisfy
	\begin{equation}\label{virtual_aod}
		\begin{aligned}
			g_{tx} = \left[\left(g_{\rm t}-1\right)\bmod G\right]+1,~~
			g_{ty} = \biggl\lceil\frac{g_{\rm t}}{G}\biggr\rceil,
		\end{aligned}
	\end{equation}
	respectively. Each column of $\tilde{{\bf U}}_{\rm t}$, i.e., $\left[\tilde{{\bf U}}_{\rm t}\right]_{:,m_{\rm d}}, m_{\rm d} = 1,\cdots, M_{\rm d}$, represents the effective channel response at the $k_{m_{\rm d}}$-th subcarrier, which exhibits the effective path sparsity \cite{10497534}. Specifically, for MA systems, the limited moving regions, i.e., $\mathcal{C}_{\rm t}$ and $\mathcal{C}_{\rm r}$, lead to a limited angular resolution. In other words, multiple Tx/Rx-MPCs with similar AoDs/AoAs can be regarded as one single effective Tx/Rx-MPC with respect to the Tx/Rx region. Moreover, since the propagation characteristics are almost unchanged within the system bandwidth, the subchannels within the system bandwidth share the same AoDs \cite{8846224,8306126}. This indicates that the columns $\left[\tilde{{\bf U}}_{\rm t}\right]_{:,m_{\rm d}}, m_{\rm d} = 1,\cdots, M_{\rm d},$ have the common sparsity, i.e., 
	\begin{equation}
		\begin{small}
			\begin{aligned}
				{\rm supp}\left\{\left[\tilde{{\bf U}}_{\rm t}\right]_{:,1}\right\} = \cdots = {\rm supp}\left\{\left[\tilde{{\bf U}}_{\rm t}\right]_{:,M_{\rm d}}\right\} = \varPsi_{\rm t},
			\end{aligned}
		\end{small}
	\end{equation}
	in which $\varPsi_{\rm t}$ is the support set. It can be concluded that there are only $L_{\rm t}$ non-zero rows in $\tilde{{\bf U}}_{\rm t}$, i.e., $\left|\varPsi_{\rm t}\right| = L_{\rm t}$, representing the number of Tx-MPCs. Due to the effective path sparsity, $L_{\rm t}\ll G^2$ is satisfied. 
	
	Based on the common sparsity, the virtual AoDs can be estimated by recovering the indices of the non-zero rows in $\tilde{{\bf U}}_{\rm t}$, which is then transformed into a sparse signal recovery problem. It is worth noting that this expression has the form of MMV-CS problem, i.e., finding a row-sparse matrix $\tilde{{\bf U}}_{\rm t}$ to minimize $\Vert {\bf V}_{\rm t} - \sqrt{p_{\rm t}}\tilde{{\bf G}}^{\rm T}\tilde{{\bf U}}_{\rm t}\Vert_{\rm F}$, in which ${\bf V}_{\rm t}$ and $\tilde{{\bf G}}^{\rm T}$ are the received pilots in step 1 at all the Tx-MA measurement positions and the measurement matrix, respectively. The SOMP algorithm can be utilized to solve this problem \cite{8765421}. The recovery of the virtual AoDs via SOMP algorithm is summarized in Algorithm \ref{alg:somp}, in which $\epsilon_0$ is a small positive parameter to guarantee the full recovery of the Tx-MPCs considering the impact of angular discretization and noise. Moreover, $I_{\rm max}$ and $\hat{L}_{\rm t}$ represent the maximum number of iterations and the number of estimated Tx-MPCs, respectively. The output of Algorithm \ref{alg:somp} is the estimated AoDs, i.e., $\left\{\hat{\varphi}_{\rm t}^{\hat{l}_{\rm t}}, \hat{\vartheta}_{\rm t}^{\hat{l}_{\rm t}}\right\}_{\hat{l}_{\rm t} = 1}^{\hat{L}_{\rm t}}$. 
	
		\begin{small}
		\begin{algorithm}[t] 
			\label{alg:somp}
			\caption{SOMP Algorithm.}
			\begin{small}
				\begin{algorithmic}[1]
				\REQUIRE $\tilde{{\bf G}}^{\rm T}$ ($\tilde{{\bf F}}^{\rm H}$ for AoA estimation, and $\tilde{\bf D}^{\rm T}$ for delay estimation), ${\bf V}_{\rm t}$ (${\bf V}_{\rm r}$ for AoA estimation, and ${\bf V}_{\rm d}$ for delay estimation), $G$ ($G_{\rm d}$ for delay estimation), $\epsilon_0$, and $I_{\rm max}$.
				\ENSURE Estimated AoDs $\left\{\hat{\varphi}_{\rm t}^{\hat{l}_{\rm t}}, \hat{\vartheta}_{\rm t}^{\hat{l}_{\rm t}}\right\}_{\hat{l}_{\rm t} = 1}^{\hat{L}_{\rm t}}$ ( estimated AoAs $\left\{\hat{\varphi}_{\rm r}^{\hat{l}_{\rm r}}, \hat{\vartheta}_{\rm r}^{\hat{l}_{\rm r}}\right\}_{\hat{l}_{\rm r} = 1}^{\hat{L}_{\rm r}}$ for AoA estimation, and estimated delays $\left\{\hat{\tau}^{\hat{l}_{\rm d}}\right\}_{\hat{l}_{\rm d} = 1}^{\hat{L}_{\rm d}}$ for delay estimation). \\
				\STATE Initialization: ${\bf V}_{\rm a}^{\left(0\right)} = {\bf V}_{\rm t}, \mathcal{A}^{\left(0\right)} = \emptyset, \tilde{{\bf U}}_{\rm a}^{\left(0\right)}={\bf 0}, \epsilon^{\left(0\right)}=1$.	
				\FOR{$i = 1:1:I_{\rm max}$}
				\STATE Find index: $i_{\rm r,max}= \mathop{\arg\max}\limits_{i_{\rm r} \notin \mathcal{A}^{\left(i-1\right)}} \sum_{i_{\rm c}=1}^{M_{\rm d}}\left|\left[\tilde{{\bf G}}^{\rm *}\right]_{i_{\rm r},:}\left[{\bf V}_{\rm a}^{\left(i-1\right)}\right]_{:,i_{\rm c}} \right|$.
				\STATE Update \begin{small}
					${{\bf q}}^{\left(i\right)}=\frac{1}{\sqrt{p_{\rm t}}}\left[\left(\left[\tilde{{\bf G}}^{\rm T}\right]_{:,\mathcal{A}^{\left(i\right)}}\right)^{\rm H}\left[\tilde{{\bf G}}^{\rm T}\right]_{:,\mathcal{A}^{\left(i\right)}}\right]^{-1}\left(\left[\tilde{{\bf G}}^{\rm T}\right]_{:,\mathcal{A}^{\left(i\right)}}\right)^{\rm H}{\bf V}_{\rm t}$
				\end{small}.
				\STATE Update ${\bf V}_{\rm a}^{\left(i\right)} = {\bf V}_{\rm t} - \sqrt{p_{\rm t}}\left[\tilde{{\bf G}}^{\rm T}\right]_{:,\mathcal{A}^{\left(i\right)}}{{\bf q}}^{\left(i\right)}$.
				\STATE Update $\epsilon^{\left(i\right)}=\frac{\Vert {\bf V}_{\rm a}^{\left(i\right)}\Vert_{\rm F}}{\Vert{\bf V}_{\rm t}\Vert_{\rm F}}$.
				\IF{$\epsilon^{\left(i-1\right)} - \epsilon^{\left(i\right)}\leq \epsilon_0$}
				\STATE Break.
				\ENDIF
				\STATE Update $\mathcal{A}^{\left(i\right)}=\mathcal{A}^{\left(i-1\right)}\cup \left\{i_{\rm r,max}\right\}$.
				\ENDFOR
				\STATE Obtain the virtual AoDs $\left\{\hat{\varphi}_{\rm t}^{\hat{l}_{\rm t}}, \hat{\vartheta}_{\rm t}^{\hat{l}_{\rm t}}\right\}_{\hat{l}_{\rm t} = 1}^{\hat{L}_{\rm t}}$ (virtual AoAs $\left\{\hat{\varphi}_{\rm r}^{\hat{l}_{\rm r}}, \hat{\vartheta}_{\rm r}^{\hat{l}_{\rm r}}\right\}_{\hat{l}_{\rm r} = 1}^{\hat{L}_{\rm r}}$, and delays $\left\{\hat{\tau}^{\hat{l}_{\rm d}}\right\}_{\hat{l}_{\rm d} = 1}^{\hat{L}_{\rm d}}$), using index set $\mathcal{A}$ and \eqref{virtual_aod}.
			\end{algorithmic}
			\end{small}
		\end{algorithm}
	\end{small}
	
	Next, we will introduce the detailed steps of Algorithm \ref{alg:somp}. In line 3, we find the row index in which the infinite-norm of matrix $\tilde{{\bf G}}^{*}{\bf V}_{\rm a}^{\left(i-1\right)}$ is located. In lines 4-6, we update the residual ${\bf V}_{\rm a}^{\left(i\right)}$ and the normalized residual $\epsilon^{\left(i\right)}$ in the $i$-th iteration, respectively. Notably, the SOMP algorithm terminates when the decrement of the normalized residual falls below the threshold $\epsilon_0$ specified in lines 7-9. In line 10, set $\mathcal{A}$ is updated for angle estimation correspondingly. Finally, in line 11, using set $\mathcal{A}$, the virtual AoDs are estimated via \eqref{virtual_aod}. 
	
	The computational complexity is analyzed as follows. In line 3, the complexities of calculating the infinite-norm of matrix $\tilde{{\bf G}}^{*}{\bf V}_{\rm a}^{\left(i-1\right)}$ and finding index $i_{\rm r,max}$ are $\mathcal{O}\left(G^2M_{\rm t}M_{\rm d}\right)$ and $\mathcal{O}\left(G^2\right)$, respectively. In line 4, the complexity of calculating ${\bf q}^{i}$ is no larger than $\mathcal{O}\left(M_{\rm t}I_{\rm max}\left(I_{\rm max}+M_{\rm d}\right)\right) = \mathcal{O}\left(I_{\rm max}M_{\rm t}M_{\rm d}\right)$, since $M_{\rm d}>I_{\rm max}$ is required to guarantee the recovery of the AoDs. In lines 5 and 6, the complexities of updating ${\bf V}_{\rm a}^{\left(i\right)}$ and $\epsilon^{\left(i\right)}$ are $\mathcal{O}\left(I_{\rm max}M_{\rm t}M_{\rm d}\right)$ and $\mathcal{O}\left(M_{\rm t}M_{\rm d}\right)$, respectively. Notably, the number of iterations is no larger than $I_{\rm max}$, and thus the total computational complexity is no larger than $\mathcal{O}\left(I_{\rm max}G^2M_{\rm t}M_{\rm d}\right)$. 
	
	For AoA estimation, the Rx-MA moves over $M_{\rm r}$ positions, denoted by ${\bf r}^{m_{\rm r}} = \left[x_{\rm r}^{m_{\rm r}}, y_{\rm r}^{m_{\rm r}}\right], m_{\rm r} = 1,\cdots, M_{\rm r}$, while the Tx-MA is fixed at reference position ${\bf t}_0$, i.e., step 2 in Fig. \ref{fig:ma_position}. Then, the received pilots at the $k_{m_{\rm d}}$-th, $1\leq m_{\rm d}\leq M_{\rm d}$, subcarrier can be represented as
	\begin{equation}
		\begin{small}
			\begin{aligned}
			& v\left({\bf t}_0, {\bf r}^{m_{\rm r}}, k_{m_{\rm d}}\right)\\ = 
			& \sqrt{p_{\rm t}}\left\{\left[\left({\bf \Sigma}\times_1{\bf f}\left({\bf r}^{m_{\rm r}}\right)^{\rm H}\right)\times_2 {\bf g}\left({\bf t}_0\right)^{\rm T}\right]\times_3 {\bf d}\left(k_{m_{\rm d}}\right)\right\}\\&+
			n^{m_{\rm r}}\left(k_{m_{\rm d}}\right)\\ \triangleq
			&\sqrt{p_{\rm t}}{\bf f}\left({\bf r}^{m_{\rm r}}\right)^{\rm H}{\bf X}_{\rm rd}{\bf d}\left(k_{m_{\rm d}}\right)+n^{m_{\rm r}}\left(k_{m_{\rm d}}\right),
		\end{aligned}
		\end{small}
	\end{equation}
	in which $n^{m_{\rm r}}\left(k_{m_{\rm d}}\right)$ represents the noise in the received pilot at the $k_{m_{\rm d}}$-th subcarrier for the $m_{\rm r}$-th Rx-MA measurement position. Matrix ${\bf X}_{\rm rd} \triangleq {\bf \Sigma}\times_2{\bf g}\left({\bf t}_0\right)^{\rm T}\in\mathbb{C}^{L_{\rm r}\times L_{\rm d}}$ is defined for notation simplicity. Collecting all the $M_{\rm d}$ pilots at the $m_{\rm r}$-th Rx-MA measurement position, i.e., ${\bf v}_{\rm r}^{m_{\rm r}} = \left[v\left({\bf t}_0, {\bf r}^{m_{\rm r}}, 0\right),\cdots, v\left({\bf t}_0, {\bf r}^{m_{\rm r}}, k_{m_{\rm d}}\right)\right]\in\mathbb{C}^{1\times M_{\rm d}}$, we have
	\begin{equation}
		{\bf v}_{\rm r}^{m_{\rm r}} = \sqrt{p_{\rm t}}{\bf f}\left({\bf r}^{m_{\rm r}}\right)^{\rm H}{\bf X}_{\rm rd}{\bf D}+{\bf n}_{\rm r}^{m_{\rm r}},
	\end{equation}
	in which ${\bf n}_{\rm r}^{m_{\rm r}}=\left[n^{m_{\rm r}}\left(0\right),\cdots,n^{m_{\rm r}}\left(k_{m_{\rm d}}\right)\right]\in\mathbb{C}^{1\times M_{\rm d}}$. Then, combining all the received pilots at all the $M_{\rm r}$ Rx-MA measurement positions in step 2, we have
	\begin{equation}\label{received_pilots_rx}
		{\bf V}_{\rm r} = \sqrt{p_{\rm t}}{\bf F}^{\rm H}{\bf X}_{\rm rd}{\bf D}+{\bf N}_{\rm r},
	\end{equation}
	in which ${\bf F} = \left[{\bf f}\left({\bf r}^{1}\right),\cdots,{\bf f}\left({\bf r}^{M_{\rm r}}\right)\right]\in\mathbb{C}^{L_{\rm r}\times M_{\rm r}}$ and ${\bf N}_{\rm r} = \left[\left({\bf n}_{\rm r}^{1}\right)^{\rm T},\cdots,\left({\bf n}_{\rm r}^{M_{\rm r}}\right)^{\rm T}\right]^{\rm T}\in\mathbb{C}^{M_{\rm r}\times M_{\rm d}}$. 
	
	To estimate the virtual AoAs via the MMV-CS method, we discretize the virtual AoAs. Specifically, we define sets $\varTheta_{\rm r}$ and $\varPhi_{\rm r}$ as
		\begin{equation}
		\begin{small}
			\left\{
		\begin{aligned}
			\varPhi_{\rm r} = \left\{\tilde{\varphi}_{\rm r}^{g_{rx}}=-1+\frac{2g_{rx}-1}{G}\left|1\leq g_{rx}\leq G\right.\right\},\\
			\varTheta_{\rm r} = \left\{\tilde{\vartheta}_{\rm r}^{g_{ry}}=-1+\frac{2g_{ry}-1}{G}\left|1\leq g_{ry}\leq G\right.\right\}.
		\end{aligned}\right.
		\end{small}
	\end{equation}
	Then, the received pilots at all the $M_{\rm r}$ Rx-MA measurement positions in step 2, i.e., \eqref{received_pilots_rx}, can be expressed as
	\begin{equation}
		{\bf V}_{\rm r} = \sqrt{p_{\rm t}}\tilde{{\bf F}}^{\rm H}\tilde{{\bf X}}_{\rm rd}{\bf D}+{\bf E}_{\rm r}+{\bf N}_{\rm r}\triangleq \sqrt{p_{\rm t}}\tilde{{\bf F}}^{\rm H}\tilde{{\bf U}}_{\rm r}+{\bf E}_{\rm r}+{\bf N}_{\rm r},
	\end{equation}
	with $\tilde{{\bf F}} = \left[\tilde{\bf f}\left({\bf r}^{1}\right)\cdots\tilde{\bf f}\left({\bf r}^{M_{\rm r}}\right)\right]\in\mathbb{C}^{G^2\times M_{\rm r}}$, $\tilde{{\bf X}}_{\rm rd}\in\mathbb{C}^{G^2\times L_{\rm d}}$, ${\bf E}_{\rm r}\in\mathbb{C}^{M_{\rm r}\times M_{\rm d}}$, and $\tilde{{\bf U}}_{\rm r} = \tilde{{\bf X}}_{\rm rd}{\bf D}\in\mathbb{C}^{G^2\times M_{\rm d}}$. Matrix ${\bf E}_{\rm r}$ represents the discretization error. Vector $\tilde{\bf f}\left({\bf r}^{m_{\rm r}}\right)\in\mathbb{C}^{G^2\times 1}, 1\leq m_{\rm r} \leq M_{\rm r},$ is defined as the discretized Rx-FRV, i.e., 
	\begin{equation}\label{discrete_frv_rx}
		\begin{small}
			\begin{aligned}
			\tilde{\bf f}\left({\bf r}^{m_{\rm r}}\right) =& \left[e^{j\frac{2\pi}{\lambda}y_{\rm r}^{m_{\rm r}}\tilde{\vartheta}_{\rm r}^{1}},\cdots,e^{j\frac{2\pi}{\lambda}y_{\rm r}^{m_{\rm r}}\tilde{\vartheta}_{\rm r}^{G_{ry}}}\right]^{\rm T}\\
			&\otimes\left[e^{j\frac{2\pi}{\lambda}x_{\rm r}^{m_{\rm r}}\tilde{\varphi}_{\rm r}^{1}},\cdots,e^{j\frac{2\pi}{\lambda}x_{\rm r}^{m_{\rm r}}\tilde{\varphi}_{\rm r}^{G_{tx}}}\right]^{\rm T}.
		\end{aligned}
		\end{small}
	\end{equation}
	Similar to the AoD estimation, the $g_{\rm r}$-th row in $\tilde{{\bf U}}_{\rm r}$ corresponds to virtual AoAs $\varphi_{\rm r}^{g_{rx}}$ and $\vartheta_{\rm r}^{g_{ry}}$, which satisfy
	\begin{equation}\label{virtual_aoa}
		\begin{aligned}
			g_{rx} = \left[\left(g_{\rm r}-1\right)\bmod G\right]+1,~~
			g_{ry} = \biggl\lceil\frac{g_{\rm r}}{G}\biggr\rceil,
		\end{aligned}
	\end{equation}
	respectively. In addition, the columns in $\tilde{{\bf U}}_{\rm r}$ exhibit the common sparsity, i.e., 
	\begin{equation}
		\begin{small}
			\begin{aligned}
			{\rm supp}\left\{\left[\tilde{{\bf U}}_{\rm r}\right]_{:,1}\right\} = \cdots = {\rm supp}\left\{\left[\tilde{{\bf U}}_{\rm r}\right]_{:,M_{\rm d}}\right\} = \varPsi_{\rm r},
		\end{aligned}
		\end{small}
	\end{equation}
	in which $\varPsi_{\rm r}$ is the support set, and $\left|\varPsi_{\rm r}\right| = L_{\rm r}\ll G^2$.
	
	The estimation of the virtual AoAs can also be regarded as a MMV-CS problem, i.e., finding a row-sparse matrix $\tilde{{\bf U}}_{\rm r}$ to minimize $\Vert {\bf V}_{\rm r} - \sqrt{p_{\rm t}}\tilde{{\bf F}}^{\rm H}\tilde{{\bf U}}_{\rm r}\Vert_{\rm F}$, in which ${\bf V}_{\rm r}$ and $\tilde{{\bf F}}^{\rm H}$ are the received pilots at all the $M_{\rm r}$ Rx-MA measurement positions in step 2 and the measurement matrix, respectively. The SOMP method in Algorithm \ref{alg:somp} can also be utilized to solve this problem, and the estimated AoAs, i.e., $\left\{\hat{\varphi}_{\rm r}^{\hat{l}_{\rm r}}, \hat{\vartheta}_{\rm r}^{\hat{l}_{\rm r}}\right\}_{\hat{l}_{\rm r} = 1}^{\hat{L}_{\rm r}}$ can be obtained, in which $\hat{L}_{\rm r}$ denotes the number of estimated Rx-MPCs.

	\subsection{Delay Estimation}\label{delay_estimation}

	In this subsection, we consider the estimation of the delays, i.e., $\left\{\tau^{l_{\rm d}}\right\}_{l_{\rm d}=1}^{L_{\rm d}}$. The delay estimation shares much similarity with the estimation of the angles, which can also be formulated as a MMV-CS problem. Specifically, since the estimation of the delays is performed in the frequency domain, which is independent of the positions of the Tx-MA and Rx-MA, the received pilots of all the three steps can be utilized for delay estimation. Notably, the received pilots in step 1 and step 2, i.e., ${\bf V}_{\rm t}$ and ${\bf V}_{\rm r}$, are obtained in \eqref{received_pilots_tx} and \eqref{received_pilots_rx}, respectively. In step 3, both the Tx-MA and Rx-MA move over $M_{\rm c}$ positions, denoted by ${\bf t}_{\rm a}^{m_{\rm c}} = \left[x_{\rm t}^{m_{\rm c}}, y_{\rm t}^{m_{\rm c}}\right]$ and ${\bf r}_{\rm a}^{m_{\rm c}} = \left[x_{\rm r}^{m_{\rm c}}, y_{\rm r}^{m_{\rm c}}\right], m_{\rm c} = 1,\cdots, M_{\rm c}$, respectively. Then, for the $m_{\rm c}$-th position pair in step 3, the received pilot at the $k_{m_{\rm d}}$-th subcarrier can be represented as
		\begin{equation}\label{receive_pilot_single_c}
		\begin{small}
			\begin{aligned}
			& v\left({\bf t}_{\rm a}^{m_{\rm c}}, {\bf r}_{\rm a}^{m_{\rm c}}, k_{m_{\rm d}}\right)\\ = 
			& \sqrt{p_{\rm t}}\left\{\left[\left({\bf \Sigma}\times_1{\bf f}\left({\bf r}_{\rm a}^{m_{\rm c}}\right)^{\rm H}\right)\times_2 {\bf g}\left({\bf t}_{\rm a}^{m_{\rm c}}\right)^{\rm T}\right]\times_3 {\bf d}\left(k_{m_{\rm d}}\right)^{\rm T}\right\}\\&+
			n^{m_{\rm c}}\left(k_{m_{\rm d}}\right)\\
		 \overset{\left({\rm a}\right)}{=}& \sqrt{p_{\rm t}}\left({\bf g}\left({\bf t}_{\rm a}^{m_{\rm c}}\right)^{\rm T}\otimes{\bf f}\left({\bf r}_{\rm a}^{m_{\rm c}}\right)^{\rm H}\right){\bf X}{\bf d}\left(k_{m_{\rm d}}\right)+n^{m_{\rm c}}\left(k_{m_{\rm d}}\right),
		\end{aligned}
		\end{small}
	\end{equation}
	in which $n^{m_{\rm c}}\left(k_{m_{\rm d}}\right)$ represents the noise in the received pilot at the $k_{m_{\rm d}}$-th subcarrier for the $m_{\rm c}$-th position pair in step 3. Step (a) in \eqref{receive_pilot_single_c} involves the matricization of PRT ${\bf \Sigma}$, as shown in Appendix \ref{sec:matrix}. Matrix ${\bf X} = \left[\left[\bf \Sigma\right]_{:,1,:}^{\rm T}, \cdots, \left[{\bf \Sigma}\right]_{:,L_{\rm t},:}^{\rm T}\right]^{\rm T}\in\mathbb{C}^{L_{\rm t}L_{\rm r}\times L_{\rm d}}$ with $\left[{\bf \Sigma}\right]_{:,l_{\rm t},:}, l_{\rm t} = 1,\cdots, L_{\rm t},$ is the lateral slice of PRT ${\bf \Sigma}$. Collecting all the $M_{\rm d}$ received pilots at the $m_{\rm c}$-th Tx/Rx-MA measurement position pair, i.e., ${\bf v}_{\rm c}^{m_{\rm c}} = \left[v\left({\bf t}^{m_{\rm c}},{\bf r}^{m_{\rm c}},0\right),\cdots,v\left({\bf t}^{m_{\rm c}},{\bf r}^{m_{\rm c}},k_{m_{\rm d}}\right)\right]\in\mathbb{C}^{1\times M_{\rm d}}$, we have
	\begin{equation}
		{\bf v}_{\rm c}^{m_{\rm c}}=\sqrt{p_{\rm t}}\left({\bf g}\left({\bf t}_{\rm a}^{m_{\rm c}}\right)^{\rm T}\otimes{\bf f}\left({\bf r}_{\rm a}^{m_{\rm c}}\right)^{\rm H}\right){\bf X}{\bf D}+{\bf n}_{\rm c}^{m_{\rm c}},
	\end{equation}
	in which ${\bf n}_{\rm c}^{m_{\rm c}} = \left[n^{m_{\rm c}}\left(0\right),\cdots,n^{m_{\rm c}}\left(k_{m_{\rm d}}\right)\right]\in\mathbb{C}^{1\times M_{\rm d}}$. Then, collecting all the received pilots at all the $M_{\rm c}$ Tx/Rx-MA position pairs in step 3, we have
	\begin{equation}\label{received_pilots_com}
		\begin{small}
			{\bf V}_{\rm c} = \sqrt{p_{\rm t}}\underbrace{\left[
		\begin{aligned}
			{\bf g}\left({\bf t}_{\rm a}^{1}\right)^{\rm T}&\otimes{\bf f}\left({\bf r}_{\rm a}^{1}\right)^{\rm H}\\
			\vdots\\
			{\bf g}\left({\bf t}_{\rm a}^{M_{\rm c}}\right)^{\rm T}&\otimes{\bf f}\left({\bf r}_{\rm a}^{M_{\rm c}}\right)^{\rm H}
		\end{aligned}
		\right]}_{\triangleq{\bf A}}{\bf X}{\bf D}+{\bf N}_{\rm c},
		\end{small}
	\end{equation}
	in which ${\bf A}\in\mathbb{C}^{M_{\rm c}\times L_{\rm t}L_{\rm r}}$ is defined for notation simplicity, and ${\bf N}_{\rm c} = \left[\left({\bf n}_{\rm c}^{1}\right)^{\rm T}, \cdots, \left({\bf n}_{\rm c}^{M_{\rm c}}\right)^{\rm T}\right]^{\rm T}\in\mathbb{C}^{M_{\rm c}\times M_{\rm d}}$. 
	
	We combine all the measured pilots in all the three steps under different Tx-MA and Rx-MA position pairs, i.e., ${\bf V}_{\rm t}$, ${\bf V}_{\rm r}$, and ${\bf V}_{\rm c}$, as
	\begin{equation}\label{received_pilots_all}
		\begin{aligned}
			&{\bf V}_{\rm d} = \left[{\bf V}_{\rm t}^{\rm T} , {\bf V}_{\rm r}^{\rm T} , {\bf V}_{\rm c}^{\rm T}\right]\\
			=&\sqrt{p_{\rm t}}{\bf D}^{\rm T}\underbrace{\left[{\bf U}_{\rm dt} , {\bf U}_{\rm dr} , {\bf U}_{\rm dc}\right]}_{{\bf U}_{\rm d}}+\underbrace{\left[{\bf N}_{\rm t}^{\rm T} , {\bf N}_{\rm r}^{\rm T} , {\bf N}_{\rm c}^{\rm T}\right]}_{{\bf N}_{\rm d}}\\
			\triangleq& \sqrt{p_{\rm t}}{\bf D}^{\rm T}{\bf U}_{\rm d}+{\bf N}_{\rm d}.
		\end{aligned}
	\end{equation}
	with ${\bf U}_{\rm dt} = {\bf X}_{\rm td}^{\rm T}{\bf G}$, ${\bf U}_{\rm dr} = {\bf X}_{\rm rd}^{\rm T}{\bf F}^{*}$, and ${\bf U}_{\rm dc} = {\bf X}^{\rm T}{\bf A}^{\rm T}$. 
	
	To estimate the delays, we define the set containing all the discretized delays as
	\begin{equation}
		\mathcal{D} = \left\{\tilde{\tau}^{g_{\rm d}} = \frac{\tau_{\rm max}}{2G_{\rm d}}+\frac{\tau_{\rm max}}{G_{\rm d}}\left(g_{\rm d}-1\right)\left|1\leq g_{\rm d} \leq G_{\rm d}\right.\right\}.
	\end{equation}
	Then, \eqref{received_pilots_all} can be rewritten as
	\begin{equation}\label{received_pilots_all_discrete}
		{\bf V}_{\rm d} = \sqrt{p_{\rm t}}\tilde{\bf D}^{\rm T}\tilde{{\bf U}}_{\rm d}+{\bf E}_{\rm d}+{\bf N}_{\rm d},
	\end{equation}
	with $\tilde{\bf D} = \left[\tilde{\bf d}\left(0\right), \cdots, \tilde{\bf d}\left(k_{m_{\rm d}}\right)\right]\in\mathbb{C}^{G_{\rm d}\times M_{\rm d}}$, $\tilde{\bf U}_{\rm d}\in\mathbb{C}^{G_{\rm d}\times M_{\rm a}}$, and ${\bf E}_{\rm d}\in\mathbb{C}^{M_{\rm d}\times M_{\rm a}}$, in which $M_{\rm a} = M_{\rm t}+M_{\rm r}+M_{\rm c}$ for notation simplicity. Matrix ${\bf E}_{\rm d}$ is the discretization error, representing the mismatch between the discretized delays in set $\mathcal{D}$ and continuous delays $\left\{\tau^{l_{\rm d}}\right\}_{l_{\rm d} = 1}^{L_{\rm d}}$. Vector $\tilde{\bf d}\left(k_{m_{\rm d}}\right)\in\mathbb{C}^{G_{\rm d}\times 1}, 1\leq m_{\rm d}\leq M_{\rm d},$ is the discretized DRV, i.e.,
	\begin{equation}\label{discrete_drv}
		\begin{aligned}
				\tilde{\bf d}\left(k_{m_{\rm d}}\right) = \left[e^{-j2\pi \frac{k_{m_{\rm d}} \tilde{\tau}^{1}}{KT_{\rm s}}},\cdots,e^{-j2\pi \frac{k_{m_{\rm d}} \tilde{\tau}^{G_{\rm d}}}{KT_{\rm s}}}\right]^{\rm T}.
		\end{aligned}
	\end{equation}
	
	Notably, each row index of $\tilde{\bf U}_{\rm d}$ uniquely corresponds to one delay $\tilde{\tau}^{g_{\rm d}}, 1\leq g_{\rm d}\leq G_{\rm d},$ defined in set $\mathcal{D}$. Moreover, since the estimation of the delays is independent of the movements of the MAs as well as the subcarriers, matrix $\tilde{\bf U}_{\rm d}$ satisfies
	\begin{equation}
		{\rm supp}\left\{\left[\tilde{{\bf U}}_{\rm d}\right]_{:,1}\right\} = \cdots = {\rm supp}\left\{\left[\tilde{{\bf U}}_{\rm d}\right]_{:,M_{\rm a}}\right\} = \varPsi_{\rm d},
	\end{equation}
	in which $\varPsi_{\rm d}$ is the support set that satisfies $\left|\varPsi_{\rm d}\right| = L_{\rm d}\ll G_{\rm d}$.
	
	According to \eqref{received_pilots_all_discrete}, the estimation of delays can also be regarded as a MMV-CS problem similar to the estimation of angles, i.e., finding a row-sparse matrix $\tilde{\bf U}_{\rm d}$ to minimize $\Vert {\bf V}_{\rm d} - \sqrt{p_{\rm t}}\tilde{\bf D}^{\rm T}\tilde{{\bf U}}_{\rm d}\Vert_{\rm F}$, in which $\tilde{\bf D}^{\rm T}$ is the measurement matrix. Then, using the SOMP method in Algorithm \ref{alg:somp}, we can obtain the estimated delays $\left\{\hat{\tau}^{\hat{l}_{\rm d}}\right\}_{\hat{l}_{\rm d} = 1}^{\hat{L}_{\rm d}}$, where $\hat{L}_{\rm d}$ is the number of estimated delays.
	
	\subsection{PRT Estimation}\label{sec:prt_est}
	
	In this subsection, we consider the estimation of the PRT via the LS method based on the estimated AoDs, AoAs, and delays. Notably, the received pilots of all the three steps are utilized to guarantee the PRT recovery and channel estimation performance, which will be shown in Section \ref{section_5}.
	
	
	To facilitate the PRT estimation, we rewrite the received pilots in steps 1 and 2 as
	\begin{equation}\label{received_pilots_txrx}
		\begin{aligned}
			&{\bf V}_{\rm t} = \sqrt{p_{\rm t}}\left[{\bf G}^{\rm T}\otimes {\bf f}\left({\bf r}_0\right)^{\rm H}\right]{\bf X}{\bf D}+{\bf N}_{\rm t},\\
			&{\bf V}_{\rm r} = \sqrt{p_{\rm t}}\left[{\bf g}\left({\bf t}_0\right)^{\rm T}\otimes {\bf F}^{\rm H}\right]{\bf X}{\bf D}+{\bf N}_{\rm r},
		\end{aligned}
	\end{equation}
	respectively, in which ${\bf X}$ is the matricization of PRT ${\bf \Sigma}$ as defined in \eqref{receive_pilot_single_c}. We combine all the $M_{\rm a}$ received pilots in all the three steps, i.e., \eqref{received_pilots_txrx} and \eqref{received_pilots_com}, into a matrix form as
	\begin{equation}\label{received_pilots_all2}
		\begin{aligned}
			{\bf V}_{\rm d}^{\rm T} &= \left[\begin{aligned}
			{\bf V}_{\rm t}\\ {\bf V}_{\rm r}\\ {\bf V}_{\rm c}
		\end{aligned}\right] = \underbrace{\left[\begin{aligned}
		{\bf G}^{\rm T}\otimes &{\bf f}\left({\bf r}_0\right)^{\rm H}\\{\bf g}\left({\bf t}_0\right)^{\rm T}&\otimes {\bf F}^{\rm H}\\ &{\bf A}
		\end{aligned}\right]}_{\triangleq{\bf \Psi}}{\bf XD}+\left[\begin{aligned}
		{\bf N}_{\rm t}\\ {\bf N}_{\rm r}\\ {\bf N}_{\rm c}
		\end{aligned}\right]\\
		&\triangleq {\bf \Psi}{\bf X}{\bf D}+{\bf N}_{\rm d}^{\rm T},
		\end{aligned}
	\end{equation}
	with ${\bf \Psi}\in \mathbb{C}^{M_{\rm a}\times L_{\rm t}L_{\rm r}}$. Note that matrices ${\bf \Psi}$ and ${\bf D}$ are determined by actual angles $\left\{\varphi_{\rm t}^{l_{\rm t}}, \vartheta_{\rm t}^{l_{\rm t}}\right\}_{l_{\rm t} = 1}^{L_{\rm t}}, \left\{\varphi_{\rm r}^{l_{\rm r}}, \vartheta_{\rm r}^{l_{\rm r}}\right\}_{l_{\rm r} = 1}^{L_{\rm r}}$ and actual delays $\left\{\tau^{l_{\rm d}}\right\}_{l_{\rm d} = 1}^{L_{\rm d}}$, which are unknown in practice.  Based on the estimated AoDs, AoAs, and delays, matrix ${\bf X}$ can be estimated by the LS method as
	\begin{equation}\label{PRT_estimation}
		{\hat {\bf X}} = {\hat {\bf \Psi}}^{\dagger}{\bf V}_{\rm d}^{\rm T}{\hat {\bf D}}^{\dagger},
	\end{equation}
	with ${\hat{\bf D}}\in\mathbb{C}^{\hat{L}_{\rm d}\times M_{\rm d}}$ and $\hat{\bf \Psi}\in\mathbb{C}^{M_{\rm a}\times \hat{L}_{\rm t}\hat{L}_{\rm r}}$. Measurement matrices ${\hat{\bf \Psi}}$ and $\hat{\bf D}$ have the same structures as ${\bf \Psi}$ and ${\bf D}$, while they are constructed by the estimated angles $\left\{\hat{\varphi}_{\rm t}^{\hat{l}_{\rm t}}, \hat{\vartheta}_{\rm t}^{\hat{l}_{\rm t}}\right\}_{\hat{l}_{\rm t} = 1}^{\hat{L}_{\rm t}}, \left\{\hat{\varphi}_{\rm r}^{\hat{l}_{\rm r}}, \hat{\vartheta}_{\rm r}^{\hat{l}_{\rm r}}\right\}_{\hat{l}_{\rm r} = 1}^{\hat{L}_{\rm r}}$, and delays $\left\{\hat{\tau}^{l_{\rm d}}\right\}_{\hat{l}_{\rm d} = 1}^{\hat{L}_{\rm d}}$, which have been obtained in steps 1-3 as specified in Section \ref{angle_estimation} and Section \ref{delay_estimation}.  
	
	\subsection{Overall Algorithm}
	
	Finally, the estimated CFR at the $k$-th subcarrier, $0\leq k\leq K-1$, can be obtained as
	\begin{equation}\label{estimate_cfr}
		\hat{h}\left({\bf t,r},k\right) = \left[\left(\hat{\bf \Sigma}\right)\times_1 \hat{\bf f}\left({\bf r}\right)^{\rm H}\times_2 \hat{\bf g}\left({\bf t}\right)^{\rm T}\right]\times_3 \hat{{\bf d}}\left(k\right)^{\rm T},
	\end{equation}
	in which the estimated PRT $\hat{\bf \Sigma}\in\mathbb{C}^{\hat{L}_{\rm r}\times\hat{L}_{\rm t}\times\hat{L}_{\rm d}}$ is obtained by reshaping $\hat{\bf X}$ to a tensor, i.e., the $l_{\rm a}$-th row $l_{\rm d}$-th column element in $\hat{\bf X}$ corresponds to the $\left(\left[\left(l_{\rm a}-1\right)\bmod \hat{L}_{\rm r}\right]+1\right)$-th row, $\lceil l_{\rm a}/\hat{L}_{\rm r}\rceil$-th column, and $l_{\rm d}$-th tube element in $\hat{\bf \Sigma}$ with $\hat{l} = \left[\left(l-1\right)\bmod \hat{L}_{\rm t}\hat{L}_{\rm r}\right]+1$. Vectors $\hat{\bf g}\left({\bf t}\right)\in\mathbb{C}^{\hat{L}_{\rm t}\times 1}$, $\hat{\bf f}\left({\bf r}\right)\in\mathbb{C}^{\hat{L}_{\rm r}\times 1}$, and $\hat{{\bf d}}\left(k\right)\in\mathbb{C}^{\hat{L}_{\rm d}\times 1}$ are estimated Tx-FRV, Rx-FRV, and DRV, with \begin{small}
		$\left[\hat{\bf g}\left({\bf t}\right)\right]_{\hat{l}_{\rm t}} = e^{j\frac{2\pi}{\lambda}\left(x_{\rm t}\hat{\varphi}_{\rm t}^{\hat{l}_{\rm t}}+y_{\rm t}\hat{\vartheta_{\rm t}}^{\hat{l}_{\rm t}}\right)}
	$
	\end{small}, \begin{small}
	$\left[\hat{\bf f}\left({\bf r}\right)\right]_{\hat{l}_{\rm r}} = e^{j\frac{2\pi}{\lambda}\left(x_{\rm r}\hat{\varphi}_{\rm r}^{\hat{l}_{\rm r}}+y_{\rm r}\hat{\vartheta_{\rm r}}^{\hat{l}_{\rm r}}\right)}$
	\end{small}, and \begin{small}
	$\left[\hat{{\bf d}}\left(k\right)\right]_{\hat{l}_{\rm d}} = e^{j2\pi \frac{k\hat{\tau}^{\hat{l}_{\rm d}}}{KT_{\rm s}}}$
	\end{small}, respectively. 
	
	\begin{algorithm}[t] 
			\label{alg:openloop}
			\caption{SOMP-based channel estimation method.}
			\begin{small}
				\begin{algorithmic}[1]
				\REQUIRE ${\bf V}_{\rm t}, {\bf V}_{\rm r}, {\bf V}_{\rm c}$, $G$, $G_{\rm d}$, $\tau_{\rm max}$, $\epsilon_0$, $I_{\rm max}$.
				\ENSURE $\hat{h}\left({\bf t,r},k\right)$, $\left\{\varphi_{\rm t}^{\hat{l}_{\rm t}}, \vartheta_{\rm t}^{\hat{l}_{\rm t}}\right\}, \hat{l}_{\rm t} = 1,\cdots, \hat{L}_{\rm t}$, $\left\{\varphi_{\rm r}^{\hat{l}_{\rm r}}, \vartheta_{\rm r}^{\hat{l}_{\rm r}}\right\}, \hat{l}_{\rm r} = 1,\cdots, \hat{L}_{\rm r}$, $\hat{\tau}^{\hat{l}_{\rm d}}, \hat{l}_{\rm d} = 1,\cdots, \hat{L}_{\rm d}$, $\hat{\bf X}$. \\
				\STATE Construct measurement matrices ${\bf G}^{\rm T}$, ${\bf F}^{\rm H}$, and ${\bf D}^{\rm T}$ using \eqref{discrete_frv_tx}, \eqref{discrete_frv_rx}, and \eqref{discrete_drv}, respectively. 
				\STATE Obtain the estimated AoDs $\left\{\varphi_{\rm t}^{\hat{l}_{\rm t}}, \vartheta_{\rm t}^{\hat{l}_{\rm t}}\right\}, \hat{l}_{\rm t} = 1,\cdots, \hat{L}_{\rm t}$, AoAs $\left\{\varphi_{\rm r}^{\hat{l}_{\rm r}}, \vartheta_{\rm r}^{\hat{l}_{\rm r}}\right\}, \hat{l}_{\rm r} = 1,\cdots, \hat{L}_{\rm r}$, and delays $\hat{\tau}^{\hat{l}_{\rm d}}, \hat{l}_{\rm d} = 1,\cdots, \hat{L}_{\rm d}$ using the SOMP method, i.e., Algorithm \ref{alg:somp}.
				\STATE Construct measurement matrices $\hat{\bf \Psi}$ and $\hat{\bf D}$ based on the estimated AoDs, AoAs, and delays. 
				\STATE Obtain $\hat{\bf X}$ via \eqref{PRT_estimation}.
				\STATE Obtain $\hat{h}\left({\bf t,r},k\right)$ via \eqref{estimate_cfr}. 
			\end{algorithmic}
			\end{small}
		\end{algorithm}
	
	The detailed steps of the proposed SOMP-based channel estimation method are summarized in Algorithm \ref{alg:openloop}. Specifically, in lines 1 and 2, we estimate the AoDs, AoAs, and delays based on the methods proposed in Section \ref{angle_estimation} and \ref{delay_estimation}, respectively. Then, in lines 3 and 4, based on the estimated AoDs, AoAs, and delays, we obtain the matricized PRT $\hat{\bf X}$ via the LS estimation in \eqref{PRT_estimation}. Finally, with the recovered wideband MPC information, the CFR between any position pair within the Tx region and the Rx region at the $k$-th, $0\leq k \leq K-1$, subcarrier can be reconstructed via the field-response channel model in \eqref{estimate_cfr}. 
	
	Next, we analyze the computational complexity of Algorithm \ref{alg:openloop}. According to the analysis in Section \ref{angle_estimation}, the computational complexity of estimating the AoDs, AoAs, and delays are no larger than $\mathcal{O}\left(I_{\rm max}G^2M_{\rm t}M_{\rm d}\right)$, $\mathcal{O}\left(I_{\rm max}G^2M_{\rm r}M_{\rm d}\right)$, and $\mathcal{O}\left(I_{\rm max}G_{\rm d}M_{\rm d}M_{\rm a}\right)$, respectively. Thus, the complexity of line 2 is $\mathcal{O}\left(I_{\rm max}M_{\rm d}\left[G^2\left(M_{\rm t}+M_{\rm r}\right)+G_{\rm d}M_{\rm a}\right]\right)$. In line 4, the computational complexity for calculating $\hat{{\bf X}}$ via the LS method is $\mathcal{O}\left(I_{\rm max}^2M_{\rm a}^2M_{\rm d}\right)$. Finally, the total computational complexity of the SOMP-based channel estimation method is $\mathcal{O}\left(I_{\rm max}G^2M_{\rm t}M_{\rm d}+ I_{\rm max}G^2M_{\rm r}M_{\rm d}+I_{\rm max}G_{\rm d}M_{\rm a}M_{\rm d}+\right.$\\$\left.I_{\rm max}^{2}M_{\rm a}^{2}M_{\rm d}\right)$.

	\section{Alternating Refinement}\label{section_4}
	
	For the proposed SOMP-based channel estimation method, the estimated wideband CSI may not be accurate enough due to the discretization errors and noise. Thus, in this section, we propose an alternating refinement method to further enhance the precision of the estimated FRVs, DRV, and PRT by minimizing the discrepancy between the received pilots and those reconstructed by the estimated CSI via the gradient descent algorithm.
	
	According to \eqref{received_pilots_all2}, all the received pilots in all the three steps can be decomposed into matrices ${\bf \Psi}$, ${\bf X}$, and ${\bf D}$. These matrices contain all the wideband MPC information, which is unknown and to be estimated. Denote ${\bf a}\in\mathbb{C}^{2\left(L_{\rm t}+L_{\rm r}\right)\times 1}$ and ${\bf b}\in\mathbb{C}^{L_{\rm d}\times 1}$ as the vectors of the angles and delays, i.e., 
	\begin{equation}\label{est_vec}
		\begin{small}
			\left\{
			\begin{aligned}
				&{\bf a} = \left[\varphi_{\rm t}^{1},\cdots, \varphi_{\rm t}^{{L}_{\rm t}}, \vartheta_{\rm t}^{1},\cdots, \vartheta_{\rm t}^{{L}_{\rm t}}, {\varphi}_{\rm r}^{1},\cdots, {\varphi}_{\rm r}^{{L}_{\rm r}}, {\vartheta}_{\rm r}^{1},\cdots, {\vartheta}_{\rm r}^{{L}_{\rm r}}\right]^{\rm T},\\
				&{{\bf b}} = \left[{\tau}^1,\cdots, {\tau}^{{L}_{\rm d}}\right]^{\rm T},
			\end{aligned}
			\right.
		\end{small}
	\end{equation}
	respectively. In this regard, the angle recovery and delay recovery are to estimate vectors ${\bf a}$ and ${\bf b}$, which are then utilized to construct matrices ${\bf \Psi}$ and ${\bf D}$, respectively. In other words, ${\bf \Psi}$ and ${\bf D}$ are the functions of vectors ${\bf a}$ and ${\bf b}$, respectively. Moreover, by recovering ${\bf X}$, the channel gains of the corresponding propagation paths can be obtained. 
	
	To this end, the optimization problem for estimating ${\bf a}$, ${\bf b}$, and ${\bf X}$ can be formulated as
	\begin{subequations}
		\begin{align}
			\label{con:obj_function}\mathop{\rm min}_{{\bf a}, {\bf b}, {\bf X}}~~&\frac{\Vert {\bf V}_{\rm d}^{\rm T} - {\bf \Psi}{{\bf X}}{{\bf D}}\Vert_{\rm F}^{2}}{\Vert {\bf V}_{\rm d}^{\rm T}\Vert_{\rm F}^{2}}\\
			\label{con:angle1}{\rm s.t.}~~& -1\leq {\varphi}_{\rm t}^{{l}_{\rm t}} \leq 1, {l}_{\rm t} = 1,\cdots, {L}_{\rm t}, \\
			\label{con:angle2}& -1\leq {\vartheta}_{\rm t}^{{l}_{\rm t}} \leq 1, {l}_{\rm t} = 1,\cdots, {L}_{\rm t}, \\
			\label{con:angle3}& -1\leq {\varphi}_{\rm r}^{{l}_{\rm r}} \leq 1, {l}_{\rm r} = 1,\cdots, {L}_{\rm r}, \\
			\label{con:angle4}& -1\leq {\vartheta}_{\rm r}^{{l}_{\rm r}} \leq 1, {l}_{\rm r} = 1,\cdots, {L}_{\rm r}, \\
			\label{con:delay}&~~ 0\leq {\tau}^{{l}_{\rm d}}\leq \tau_{\rm max}, {l}_{\rm d} = 1,\cdots, {L}_{\rm d},
		\end{align}%
	\end{subequations}
	in which \eqref{con:obj_function} represents the square of the discrepancy between the received pilots and those reconstructed by the estimated CSI. Constraints \eqref{con:angle1}, \eqref{con:angle2}, \eqref{con:angle3}, and \eqref{con:angle4} confine that the estimated angles are within $\left[-1,1\right]$. Constraint \eqref{con:delay} indicates that the estimated delays should be no larger than the maximum delay.

	To solve the optimization problem, first, for any given ${\bf \Psi}$ and ${\bf D}$, matrix ${\bf X}$ can be estimated via the LS method in closed-form in \eqref{PRT_estimation}, i.e., $\hat{{\bf X}} = {\bf \Psi}^{\dagger}{\bf V}_{\rm d}^{\rm T}{\bf D}^{\dagger}$. Next, we focus on the estimation of the AoDs, AoAs, and delays, i.e., vectors ${\bf a}$ and ${\bf b}$. To this end, the gradient descent method is employed to estimate the AoDs, AoAs, and delays. Accordingly, the optimization problem becomes
	\begin{subequations}\label{opt_pro}
		\begin{align}
			\label{obj_func}\mathop{\rm min}_{{\bf a}, {\bf b}}~~&\frac{\Vert {\bf V}_{\rm d}^{\rm T} - {\bf \Psi}{\bf \Psi}^{\dagger}{\bf V}_{\rm d}^{\rm T}{\bf D}^{\dagger}{{\bf D}}\Vert_{\rm F}^{2}}{\Vert {\bf V}_{\rm d}^{\rm T}\Vert_{\rm F}^{2}} \triangleq g\left({\bf a}, {\bf b}\right),\\
			{\rm s.t.}~~&\eqref{con:angle1}, \eqref{con:angle2}, \eqref{con:angle3}, \eqref{con:angle4}, {\rm and}~ \eqref{con:delay}.
		\end{align}%
	\end{subequations}
	Since the initial values are critical for solving the optimization problem \eqref{opt_pro}, the initial vectors ${\bf a}^{\left(0\right)}\in\mathbb{C}^{2\left(\hat{L}_{\rm t}+\hat{L}_{\rm r}\right)\times 1}$ and ${\bf b}^{\left(0\right)}\in\mathbb{C}^{\hat{L}_{\rm d}\times 1}$ are the estimated angles and delays in the SOMP-based channel estimation in Section \ref{section_3}, with $\hat{L}_{\rm t}, \hat{L}_{\rm r}$, and $\hat{L}_{\rm d}$ denoting the numbers of the estimated Tx-MPCs, Rx-MPCs, and delays, respectively. Then, based on the initial vectors, the angles and delays are refined iteratively via the gradient descent algorithm. Denote ${\bf a}^{\left(i-1\right)}$ and ${\bf b}^{\left(i-1\right)}$ as the matrices obtained in the $\left(i-1\right)$-th iteration. The gradient of function $g\left({\bf a}, {\bf b}\right)$ w.r.t. ${\bf a}$ and ${\bf b}$ at ${\bf a}^{\left(i-1\right)}$ and ${\bf b}^{\left(i-1\right)}$ can be calculated according to the definition, i.e., 
	\begin{equation}\label{gradient}
		\begin{small}
			\left\{
		\begin{aligned}
			&\left[\nabla_{\bf a}g\left({\bf a}^{\left(i-1\right)}, {\bf b}^{\left(i-1\right)}\right)\right]=\frac{\partial g\left({\bf a}, {\bf b}\right)}{\partial\left[{\bf a}\right]_{i_{\rm a}}}\Bigg|_{{\bf a} = {\bf a}^{\left(i-1\right)},{\bf b} = {\bf b}^{\left(i-1\right)}} \\
			=&\lim\limits_{\delta\rightarrow 0}\frac{g\left({\bf a}^{\left(i-1\right)}+\delta {\bf z}_{2\left(\hat{L}_{\rm t}+\hat{L}_{\rm r}\right)}^{i_{\rm a}},{\bf b}^{\left(i-1\right)}\right) - g\left({\bf a}^{\left(i-1\right)},{\bf b}^{\left(i-1\right)}\right)}{\delta},\\
			&i_{\rm a} = 1,\cdots, 2\left(\hat{L}_{\rm t}+\hat{L}_{\rm r}\right),\\
			&\left[\nabla_{{\bf b}}g\left({\bf a}^{\left(i-1\right)}, {\bf b}^{\left(i-1\right)}\right)\right]=\frac{\partial g\left({\bf a}, {\bf b}\right)}{\partial\left[{\bf b}\right]_{i_{\rm b}}}\Bigg|_{{\bf a} = {\bf a}^{\left(i-1\right)},{\bf b} = {\bf b}^{\left(i-1\right)}} \\
			=&\lim\limits_{\delta\rightarrow 0}\frac{g\left({\bf a}^{\left(i-1\right)},{\bf b}^{\left(i-1\right)}+\delta {\bf z}_{\hat{L}_{\rm d}}^{i_{\rm b}}\right) - g\left({\bf a}^{\left(i-1\right)},{\bf b}^{\left(i-1\right)}\right)}{\delta},\\
			&i_{\rm b} = 1,\cdots, \hat{L}_{\rm d},
		\end{aligned}\right.
		\end{small}
	\end{equation}
	in which ${\bf z}_{2\left(\hat{L}_{\rm t}+\hat{L}_{\rm r}\right)}^{i_{\rm a}}$ is a $2\left(\hat{L}_{\rm t}+\hat{L}_{\rm r}\right)$-dimensional vector with a one at the $i_{\rm a}$-th element and zeros elsewhere. Similarly, ${\bf z}_{\hat{L}_{\rm d}}^{i_{\rm b}}$ is defined as the $\hat{L}_{\rm d}$-dimensional vector with a one at the $i_{\rm b}$-th element and zeros elsewhere. Then, the estimated angle and delay vectors in the $i$-th iteration are updated as
	\begin{equation}\label{vector_update}
		\left\{
		\begin{aligned}
			&{\bf a}^{\left(i\right)} = \mathcal{B}\left\{{\bf a}^{\left(i-1\right)} - \delta_{\rm a}^{\left(i\right)}\nabla_{{\bf a}}g\left({\bf a}^{\left(i-1\right)}, {\bf b}^{\left(i-1\right)}\right)\right\},\\
			&{\bf b}^{\left(i\right)} = \mathcal{B}\left\{{\bf b}^{\left(i-1\right)} - \delta_{\rm d}^{\left(i\right)}\nabla_{{\bf b}}g\left({\bf a}^{\left(i-1\right)}, {\bf b}^{\left(i-1\right)}\right)\right\},
		\end{aligned}
		\right.
	\end{equation}
	in which $\delta_{\rm a}^{\left(i\right)}$ and $\delta_{\rm d}^{\left(i\right)}$ are step sizes for the refinements of angles and delays in the $i$-th iteration, respectively. Notably, due to the large difference in the gradients of function $g\left({\bf a}, {\bf b}\right)$ w.r.t. ${\bf a}$ and ${\bf b}$, different step sizes, i.e., $\delta_{\rm a}^{\left(i\right)}$ and $\delta_{\rm d}^{\left(i\right)}$, are utilized to guarantee that both angles and delays can be refined for the $i$-th iteration. Moreover, $\mathcal{B}\left\{\cdot\right\}$ is the boundary function, which guarantees that each element in ${\bf a}$ and ${\bf b}$ remains in the feasible region, i.e.,
	\begin{equation}
		\left[\mathcal{B}\left\{{\bf a}\right\}\right]_{i} = \left\{
		\begin{aligned}
			-1,~~& {\rm if} ~\left[{\bf a}\right]_i < -1,\\
			\left[{\bf a}\right]_i,~~& {\rm if}~ -1<\left[{\bf a}\right]_i<1,\\
			1,~~& {\rm if} ~\left[{\bf a}\right]_i>1,
		\end{aligned}\right.
	\end{equation}
	and
	\begin{equation}
		\left[\mathcal{B}\left\{{\bf b}\right\}\right]_{i} = \left\{
		\begin{aligned}
			0,~~& {\rm if} ~\left[{\bf b}\right]_i < 0,\\
			\left[{\bf b}\right]_i,~~& {\rm if}~ 0<\left[{\bf b}\right]_i<\tau_{\rm max},\\
			\tau_{\rm max},~~& {\rm if} ~\left[{\bf b}\right]_i>\tau_{\rm max},
		\end{aligned}\right.
	\end{equation}
	respectively. 
	
	To obtain appropriate step sizes, backtracking line search is utilized \cite{10354003}. In particular, for each iteration, the initial step sizes are set to relatively large values, i.e., $\delta_{\rm a}^{\left(i\right)} = \delta_{\rm a}^{0}$ and $\delta_{\rm d}^{\left(i\right)} = \delta_{\rm d}^{0}$. Then, multiply the step sizes by a constant value $\kappa\in\left(0,1\right)$ to shrink them, i.e., $\delta_{\rm a}^{\left(i\right)}\leftarrow \kappa \delta_{\rm a}^{\left(i\right)}$ and $\delta_{\rm d}^{\left(i\right)}\leftarrow \kappa \delta_{\rm d}^{\left(i\right)}$, until the Armijo–Goldstein condition is satisfied, i.e,
	\begin{equation}\label{AG_condition}
		\begin{aligned}
			g\left({\bf a}^{\left(i\right)},{\bf b}^{\left(i\right)}\right)\leq& g\left({\bf a}^{\left(i-1\right)},{\bf b}^{\left(i-1\right)}\right) \\
			&- \xi \delta_{\rm a}^{\left(i\right)}\Vert \nabla_{{\bf a}}g\left({\bf a}^{\left(i-1\right)},{\bf b}^{\left(i-1\right)}\right)\Vert_2^2\\
			&-\xi \delta_{\rm d}^{\left(i\right)}\Vert\nabla_{{\bf b}} g\left({\bf a}^{\left(i-1\right)},{\bf b}^{\left(i-1\right)}\right)\Vert_2^2.
		\end{aligned}
	\end{equation} 
	
	Finally, through the alternating refinement method, the AoDs, AoAs, delays, and PRT (matricized) can be effectively estimated, denoted by $\left\{\hat{\varphi}_{\rm t}^{\hat{l}_{\rm t}}, \hat{\vartheta}_{\rm t}^{\hat{l}_{\rm t}}\right\}_{\hat{l}_{\rm t}}^{\hat{L}_{\rm t}}$, $\left\{\hat{\varphi}_{\rm r}^{\hat{l}_{\rm r}}, \hat{\vartheta}_{\rm r}^{\hat{l}_{\rm r}}\right\}_{\hat{l}_{\rm r}}^{\hat{L}_{\rm r}}$, $\left\{\hat{\tau}^{\hat{l}_{\rm d}}\right\}_{\hat{l}_{\rm d}}^{\hat{L}_{\rm d}}$, and $\hat{\bf X}$, respectively. Accordingly, the CFR can be estimated via \eqref{estimate_cfr}, in which the estimated Tx-FRV, Rx-FRV, DRV, and PRT are constructed based on the estimated wideband MPC information.
	
		\begin{small}
		\begin{algorithm}[t] 
			\label{alg:gradient_descent}
			\caption{Alternating refinement method.}
			\begin{small}
				\begin{algorithmic}[1]
				\REQUIRE Initial vectors ${\bf a}^{\left(0\right)}$ and ${\bf b}^{\left(0\right)}$ according to the SOMP-based channel estimation method, $\delta_{\rm a}^{0}$, $\delta_{\rm d}^{0}$, $\delta_{\rm min}$, $I_{\rm gd}$, $\xi$, $\kappa$, $\tau_{\rm max}$. 
				\ENSURE $\hat{h}\left({\bf t,r},k\right)$. \\
				\STATE Initialize $g\left({\bf a}^{\left(0\right)}, {\bf b}^{\left(0\right)}\right)$ via \eqref{obj_func}. 
				\FOR{$i = 1:1:I_{{\rm gd}}$}
				\STATE Calculate the gradient values $\nabla_{{\bf a}}g\left({\bf a}^{\left(i-1\right)}, {\bf b}^{\left(i-1\right)}\right)$ and $\nabla_{{\bf b}}g\left({\bf a}^{\left(i-1\right)}, {\bf b}^{\left(i-1\right)}\right)$ according to \eqref{gradient}. 
				\WHILE{$\delta_{\rm a}^{\left(i\right)}>\delta_{\rm min}$ and $\delta_{\rm d}^{\left(i\right)}>\delta_{\rm min}$}
				\STATE Update ${\bf a}^{\left(i\right)}$ and ${\bf b}^{\left(i\right)}$ via \eqref{vector_update}.  
				\STATE Construct measurement matrices ${\bf \Psi}^{\left(i\right)}$ and ${\bf D}^{\left(i\right)}$ using ${\bf a}^{\left(i\right)}$ and ${\bf b}^{\left(i\right)}$, respectively. 
				\STATE Update ${\bf X}^{\left(i\right)}$ via the LS estimation in \eqref{PRT_estimation}. 
				\IF{Condition \eqref{AG_condition} is not satisfied}
					\STATE Update step sizes $\delta_{\rm a}^{\left(i\right)}\leftarrow \kappa\delta_{\rm a}^{\left(i\right)}$ and $\delta_{\rm d}^{\left(i\right)}\leftarrow \kappa\delta_{\rm d}^{\left(i\right)}$.
				\ELSE
					\STATE break.
				\ENDIF
				\ENDWHILE
				\ENDFOR
				\STATE Obtain the estimated AoDs $\left\{\hat{\varphi}_{\rm t}^{\hat{l}_{\rm t}}, \hat{\vartheta}_{\rm t}^{\hat{l}_{\rm t}}\right\}_{\hat{l}_{\rm t}}^{\hat{L}_{\rm t}}$, AoAs $\left\{\hat{\varphi}_{\rm r}^{\hat{l}_{\rm r}}, \hat{\vartheta}_{\rm r}^{\hat{l}_{\rm r}}\right\}_{\hat{l}_{\rm r}}^{\hat{L}_{\rm r}}$, delays $\left\{\hat{\tau}^{\hat{l}_{\rm d}}\right\}_{\hat{l}_{\rm d}}^{\hat{L}_{\rm d}}$, and matricized PRT $\hat{\bf X}$. 
				\STATE Obtain $\hat{h}\left({\bf t,r}, k\right)$ via \eqref{estimate_cfr}. 
			\end{algorithmic}
			\end{small}
		\end{algorithm}
	\end{small}
	
	The detailed steps of the gradient descent-based alternating refinement are summarized in Algorithm \ref{alg:gradient_descent}, in which $\delta_{\rm min}$ represents the minimum step size. In lines 2-14, the AoDs, AoAs, and delays are refined via the iterative optimization of vectors ${\bf a}$ and ${\bf b}$ using the gradient descent method. For each iteration, appropriate step sizes are determined via the backtracking line search in lines 4-12. Moreover, in lines 6 and 7, the matricized PRT is estimated via the LS method. The overall algorithm terminates after a fixed number $I_{\rm gd}$ of iterations.  Next, we analyze the convergence of Algorithm \ref{alg:gradient_descent}. $g\left({\bf a}^{\left(i\right)}, {\bf b}^{\left(i\right)}\right)\leq g\left({\bf a}^{\left(i-1\right)}, {\bf b}^{\left(i-1\right)}\right)$ is guaranteed since $\xi>0$, $\delta_{\rm a}^{\left(i\right)}>0$, $\delta_{\rm d}^{\left(i\right)}>0$, $\Vert \nabla_{{\bf a}}g\left({\bf a}^{\left(i-1\right)},{\bf b}^{\left(i-1\right)}\right)\Vert_2^2>0$, and $\Vert \nabla_{{\bf b}}g\left({\bf a}^{\left(i-1\right)},{\bf b}^{\left(i-1\right)}\right)\Vert_2^2>0$. Since $g\left({\bf a},{\bf b}\right)$ has the form of the square of Frobenius norm, which is nonnegative, we can always find a lower bound of $g\left({\bf a},{\bf b}\right)$, yielding a suboptimal solution for problem \eqref{opt_pro}. 
	
	The computational complexity of Algorithm \ref{alg:gradient_descent} is analyzed as follows. Given that the dimensions of the initial vectors ${\bf a}^{\left(0\right)}$ and ${\bf b}^{\left(0\right)}$ are $2\left(\hat{L}_{\rm t}+\hat{L}_{\rm r}\right)$ and $\hat{L}_{\rm d}$, we have ${\bf \Psi}^{\left(0\right)}\in\mathbb{C}^{M_{\rm a}\times \hat{L}_{\rm t}\hat{L}_{\rm r}}$ and ${\bf D}^{\left(0\right)}\in\mathbb{C}^{\hat{L}_{\rm d}\times M_{\rm d}}$, respectively. In line 3, the calculations of the gradient values $\nabla_{{\bf a}}g\left({\bf a}^{\left(i-1\right)}, {\bf b}^{\left(i-1\right)}\right)$ and $\nabla_{{\bf b}}g\left({\bf a}^{\left(i-1\right)}, {\bf b}^{\left(i-1\right)}\right)$ involve the calculation of the objective function, which has the computational complexity of $\mathcal{O}\left(M_{\rm a}^2\hat{L}_{\rm t}\hat{L}_{\rm r}+M_{\rm a}^2M_{\rm d}+M_{\rm a}M_{\rm d}^2\right)$. Thus, the overall computational complexity in line 3 is $\mathcal{O}\left(\left(\hat{L}_{\rm t}+\hat{L}_{\rm r}+\hat{L}_{\rm d}\right)\left(M_{\rm a}^2\hat{L}_{\rm t}\hat{L}_{\rm r}+M_{\rm a}^2M_{\rm d}+M_{\rm a}M_{\rm d}^2\right)\right)$. Denote the maximum number of iterations for backtracking line search in lines 4-13 as $J_{\rm max}$. Then, the corresponding computational complexity is $\mathcal{O}\left(J_{\rm max}\left(\hat{L}_{\rm t}+\hat{L}_{\rm r}+\hat{L}_{\rm d}\right)\left(M_{\rm a}^2\hat{L}_{\rm t}\hat{L}_{\rm r}+M_{\rm a}^2M_{\rm d}+M_{\rm a}M_{\rm d}^2\right)\right)$. Finally, the overall complexity of Algorithm \ref{alg:gradient_descent} can be obtained as $\mathcal{O}\left(I_{\rm gd}J_{\rm max}\left(\hat{L}_{\rm t}+\hat{L}_{\rm r}+\hat{L}_{\rm d}\right)\left(M_{\rm a}^2\hat{L}_{\rm t}\hat{L}_{\rm r}+M_{\rm a}^2M_{\rm d}+\right.\right.$\\$\left.\left.M_{\rm a}M_{\rm d}^2\right)\right)$.
	
	\section{Simulation Results}\label{section_5}
	
	In this section, comprehensive simulations are carried out to evaluate the effectiveness of the proposed SOMP-based channel estimation and alternating refinement methods. 
	
	\subsection{Simulation Setup}
	
	In the simulation, the carrier frequency is set to 28 GHz with wavelength $\lambda \approx 0.0107 \rm ~m$. The system bandwidth is set to 80 MHz, which corresponds to the sampling period of $T_{\rm s} = 12.5 {\rm ns}$. The number of suncarriers is set to $K=256$. Both Tx-MA and Rx-MA can flexibly adjust their positions within the square Tx and Rx regions with normalized region size $S=3$, i.e., $\mathcal{C}_{\rm t}, \mathcal{C}_{\rm r} = \left[-\frac{3}{2}\lambda, \frac{3}{2}\lambda\right]\times\left[-\frac{3}{2}\lambda, \frac{3}{2}\lambda\right]$. For channel generation, the geometry channel model is utilized. Specifically, there are $L=6$ propagation paths, and each propagation path corresponds to one Tx-MPC and one Rx-MPC with a unique delay. In such a case, all the non-zero elements in PRT ${\bf \Sigma}$ are located on the diagonal. In other words, PRT ${\bf \Sigma}$ becomes an $\left(L\times L\times L\right)$-dimensional tensor, and only the elements with indices $\left[{\bf \Sigma}\right]_{l,l,l}, l = 1,\cdots, L,$ are non-zero. The physical AoDs and AoAs, i.e., $\left\{\theta_{\rm t}^{l}, \phi_{\rm t}^{l}\right\}_{l=1}^{L}$ and $\left\{\theta_{\rm r}^{l}, \phi_{\rm r}^{L}\right\}_{l=1}^{L}$, follow the probability density function of \cite{zhu2022modeling,10497534}
	\begin{equation}
		\begin{aligned}
			&f_{\rm AoD}\left(\theta_{\rm t}^{l}, \phi_{\rm t}^{l}\right) = \frac{\cos \theta_{\rm t}^{l}}{2\pi},~~
			&f_{\rm AoA}\left(\theta_{\rm r}^{l}, \phi_{\rm r}^{l}\right) = \frac{\cos \theta_{\rm r}^{l}}{2\pi},
		\end{aligned}
	\end{equation}
	indicating that the MPCs are uniformly distributed in front of the antenna panel. The delays of the MPCs, i.e., $\left\{\tau^{l}\right\}_{l=1}^{L}$, are uniformly distributed with $\left[0,\tau_{\rm max}\right]$, in which the maximum delay is set to $\tau_{\rm max} = 0.15~\mu s$ \cite{zhu2024performance}. The cyclic prefix (CP) length is set to $K_{\rm CP}  = 16$, which guarantees $K_{\rm CP}T_{\rm s} = 16 \times 0.0125 =~0.2 \mu s > \tau_{\rm max} = 0.15 \mu s$. Moreover, complex coefficients are independent and identically distributed (i.i.d.) circularly symmetric complex Gaussian (CSCG) random variables, i.e., $\left[\bf \Sigma\right]_{l,l,l}\in\mathcal{CN}\left(0,\frac{1}{L}\right), l = 1,\cdots, L$.
	
	We set the threshold and maximum number of iterations of the SOMP algorithm as $\epsilon_0 = 0.02$ and $I_{\rm max} = 10$, respectively. The numbers of grids for angle and delay discretizations are $G = G_{\rm d} = 100$. For alternating refinement, we have the step sizes for angle refinement $\delta_{\rm a}^{0} = 6/G = 0.06$, delay refinement $\delta_{\rm d}^{0} = \tau_{\rm max}/G_{\rm d} = 1.5\times 10^{-3}$, and minimum step size $\delta_{\rm min} = 10^{-15}$. Moreover, $I_{\rm gd} = 10$, $\xi = 0.6$, $\kappa = 0.5$. Unless otherwise specified, we set the average SNR as $p_{\rm t}/\sigma^2 = 20~\rm dB$ and number of pilot symbols $M_{\rm a} = 328$ ($M_{\rm t} = M_{\rm r} = 64$ in steps 1 and 2, as well as $M_{\rm c} = 200$ in step 3). Moreover, since the channel estimation performance is influenced by the MA measurement positions, we consider three MA measurement position setups for steps 1 and 2: 1) {\bf UPA-shape}: $\left\{{\bf t}^{m_{\rm t}}\right\}_{m_{\rm t} = 1}^{M_{\rm t}}$ and $\left\{{\bf r}^{m_{\rm r}}\right\}_{m_{\rm r} = 1}^{M_{\rm r}}$ traverse the entire Tx and Rx regions as the shape of UPA with spacing $S/\sqrt{M_{\rm t}}$ and $S/\sqrt{M_{\rm r}}$ between adjacent positions, respectively; 2) {\bf Edge of region}: $\left\{{\bf t}^{m_{\rm t}}\right\}_{m_{\rm t} = 1}^{M_{\rm t}}$ and $\left\{{\bf r}^{m_{\rm r}}\right\}_{m_{\rm r} = 1}^{M_{\rm r}}$ traverse the edges of the Tx and Rx regions with spacing $4S/M_{\rm t}$ and $4S/M_{\rm r}$ between adjacent positions; 3) {\bf Random}: $\left\{{\bf t}^{m_{\rm t}}\right\}_{m_{\rm t} = 1}^{M_{\rm t}}$ and $\left\{{\bf r}^{m_{\rm r}}\right\}_{m_{\rm r} = 1}^{M_{\rm r}}$ follow 2D uniform distributions in the Tx and Rx regions. For step 3, both $\left\{{\bf t}^{m_{\rm c}}\right\}_{m_{\rm c} = 1}^{M_{\rm c}}$ and $\left\{{\bf r}^{m_{\rm c}}\right\}_{m_{\rm c} = 1}^{M_{\rm c}}$ follow 2D uniform distributions in the Tx and Rx regions.
	
	To verify the accuracy and effectiveness of the proposed SOMP-based channel estimation and alternating refinement methods, two evaluation metrics, normalized mean square error (NMSE) and achievable rate, are utilized. Specifically, the NMSE is given by 
	\begin{equation}
		\begin{small}
			\begin{aligned}
				{\rm NMSE} = \mathbb{E}\left[\frac{\sum_{k=1}^{K}\Big\Vert \left[\mathcal{H}\right]_{:,:,k} - \left[\hat{\mathcal{H}}\right]_{:,:,k}\Big\Vert_{\rm F}^2}{\sum_{k=1}^{K}\Vert \left[\mathcal{H}\right]_{:,:,k}\Vert_{\rm F}^2}\right],
		\end{aligned}
		\end{small}
	\end{equation}
	in which $\mathcal{H}\in\mathbb{C}^{D^2\times D^2 \times K}$ is the tensor containing all the CFRs across the Tx and Rx regions of all the $K$ subcarriers. In particular, the Tx region is approximated by $D^2$ grids, which is the same for the Rx region. For the $k$-th frontal slice, i.e., $\left[\mathcal{H}\right]_{:,:,k}$, the $D^2$ elements represent the CFRs at the $k$-th subcarrier between any gird in the Tx region and any grid in the Rx region. Then, the $K$ frontal slices represent all the $K$ subcarriers. Tensor $\hat{\mathcal{H}}$ is the estimation of $\mathcal{H}$. In the simulation, we set $D^2 = 256$.
	
	In addition, the achievable rate is given by
	\begin{equation}
		\begin{small}
			\begin{aligned}
			R = \frac{1}{K+K_{\rm CP}}\sum_{k=1}^{K}\log_2\left(1+\frac{\left|h\left(\hat{{\bf t}}_{\rm max}, \hat{{\bf r}}_{\rm max}, k\right)\right|^2p_{\rm t}}{\sigma^2}\right),
		\end{aligned}
		\end{small}
	\end{equation}
	in which $\hat{{\bf t}}_{\rm max}$ and $\hat{{\bf r}}_{\rm max}$ are the positions with the maximum achievable rate obtained based on the estimated CSI. Due to the estimation error, $\hat{{\bf t}}_{\rm max}$ and $\hat{{\bf r}}_{\rm max}$ may not be the actual positions with the maximum achievable rate, and thus, the actual maximum achievable may not be obtained with the estimated CSI. In such a case, we set the maximum achievable rate obtained based on the perfect CSI as a benchmark to evaluate the effectiveness of the proposed channel estimation methods. In addition, we also set the achievable rate of the FPA system, with one single Tx antenna and one single Rx antenna fixed at the reference positions as a benchmark, to evaluate the performance gain of MA systems. 
		
	\subsection{Numerical Results}
	
\begin{figure}[t]
		\centering
		\includegraphics[width= 7 cm]{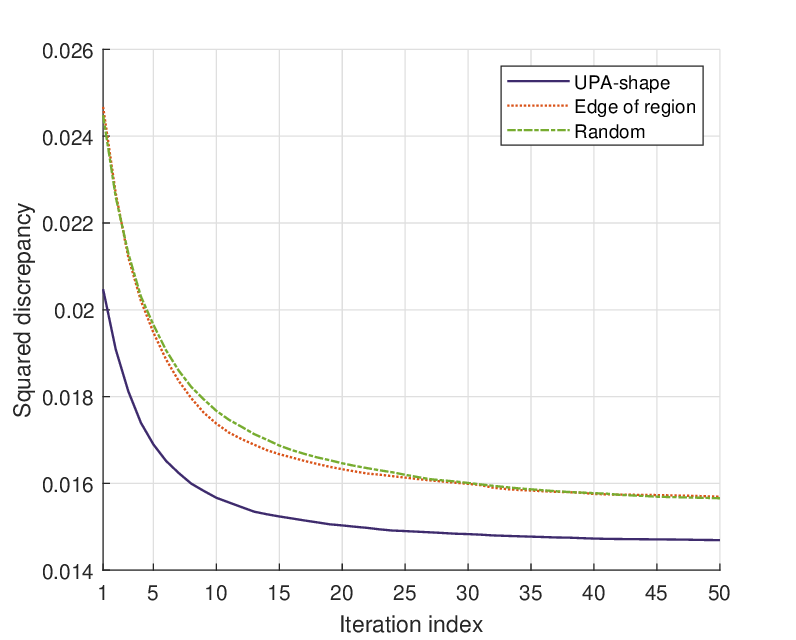}
		\caption{Evaluation of the proposed alternating refinement method under different MA measurement setups. }
		\label{fig:nmse_iter}
	\end{figure}

\begin{figure}[t]
		\centering
		\includegraphics[width= 7 cm]{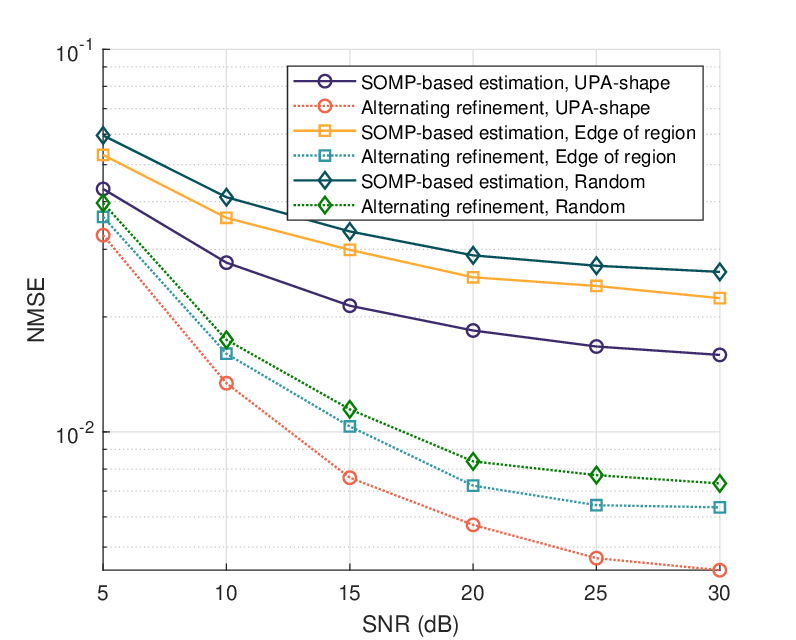}
		\caption{NMSE of the SOMP-based channel estimation method and alternating refinement method versus SNR.}
		\label{fig:snr_nmse}
	\end{figure}

	In Fig. \ref{fig:nmse_iter}, we show the convergence of the proposed alternating refinement method, i.e., Algorithm \ref{alg:gradient_descent}, under different MA measurement position setups. It can be observed that all the objective functions, i.e., square error, of the three setups decrease with the increase of iteration index, and reach a steady state after 40 iterations. In fact, through the SOMP-based channel estimation, the errors of all the three setups between estimated CFRs and received pilots are small enough to obtain the CSI with high accuracy, which will be shown in Fig. \ref{fig:snr_nmse}. The alternating refinement can further decrease the error and leads to a higher accuracy of the CSI. Moreover, the UPA-shape setup can reach a smaller error, indicating that traversing the whole regions in steps 1 and 2 can help increase the estimation accuracy.

	In Fig. \ref{fig:snr_nmse}, we evaluate the NMSE with varying SNR under different MA measurement position setups. The NMSE of the alternating refinement method can be significantly reduced under different SNR. Notably, the NMSE of the alternating refinement with 10 dB SNR can be lower than the NMSE of the SOMP-based channel estimation with 30 dB SNR. This indicates that for the same channel estimation performance, channel estimation with alternating refinement can effectively reduce the transmit power required compared to the SOMP-based channel estimation. Moreover, the channel estimation performance is also related to the MA measurement positions. Specifically, the UPA-shape setup can outperform the other two setups in terms of NMSE. This indicates that the performance can be further enhanced by traversing the entire Tx and Rx regions.  
	
\begin{figure}[t]
		\centering
		\includegraphics[width= 7 cm]{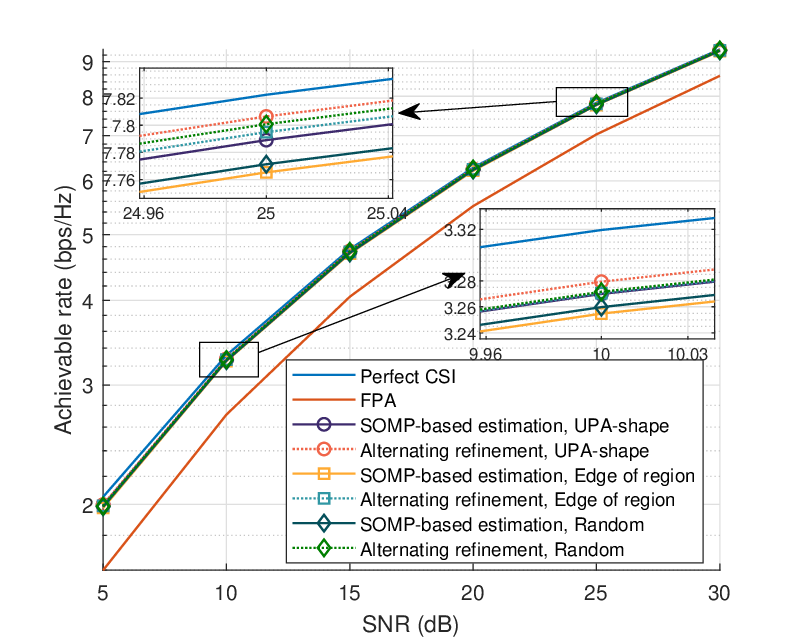}
		\caption{Achievable rates of the SOMP-based channel estimation method and alternating refinement method versus SNR.}
		\label{fig:snr_rate}
	\end{figure}

	In Fig. \ref{fig:snr_rate}, we evaluate the achievable rate with varying SNR under different MA measurement position setups. With the increase of average SNR (i.e., the transmit power $p_{\rm t}$), the achievable rates increase significantly. Compared to the FPA system, the achievable rate can be further increased by adjusting the positions of the MAs. With $5$ dB SNR, the achievable rates obtained based on the estimated CSI are lower than that based on the perfect CSI due to the large estimation error. When the SNR becomes larger than $10$ dB, both the SOMP-based channel estimation and alternating refinement methods with all the three MA measurement position setups can obtain comparable achievable rates to the perfect CSI. This indicates that sufficiently large performance gains can still be obtained with SOMP-based channel estimation method. 
	
\begin{figure}[t]
		\centering
		\includegraphics[width= 7 cm]{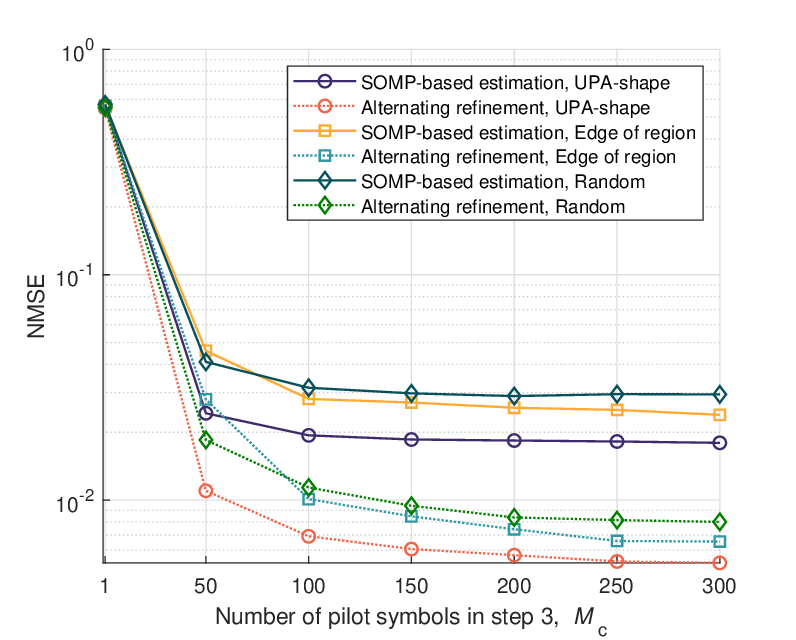}
		\caption{NMSE of the SOMP-based channel estimation method and alternating refinement method versus number $M_{\rm c}$ of pilots symbols in step 3.}
		\label{fig:m_nmse}
	\end{figure}

	The NMSE versus the number $M_{\rm c}$ of pilot symbols in step 3 is evaluated in Fig. \ref{fig:m_nmse}. In the simulation, $M_{\rm t} = M_{\rm r} = 64$. As can be observed, the channel estimation method is not implementable with a small number of $M_{\rm c}$. When $M_{\rm c}<100$, the NMSE decreases rapidly with the increase of $M_{\rm c}$. This indicates that the additional channel measurements in step 3 are necessary for both the SOMP-based channel estimation and alternating refinement methods. It can be observed that the NMSE can be further reduced with the increase of $M_{\rm c}$ when $M_{\rm c}>100$. Moreover, the NMSE of the alternating method with $M_{\rm c} = 50$ can outperform the NMSE of the SOMP-based channel estimation method with $M_{\rm c} = 100$. This indicates that the number of additional pilot symbols required can be reduced by $50\%$ by employing the alternating refinement method compared to the SOMP-based channel estimation method, which demonstrates the effectiveness of the alternating refinement method. In addition, it can still be observed that the UPA-shape setup reaches the lowest NMSE among the three setups, demonstrating the extra gains brought by the MA measurement position setups. 
	
\begin{figure}[t]
		\centering
		\includegraphics[width= 7 cm]{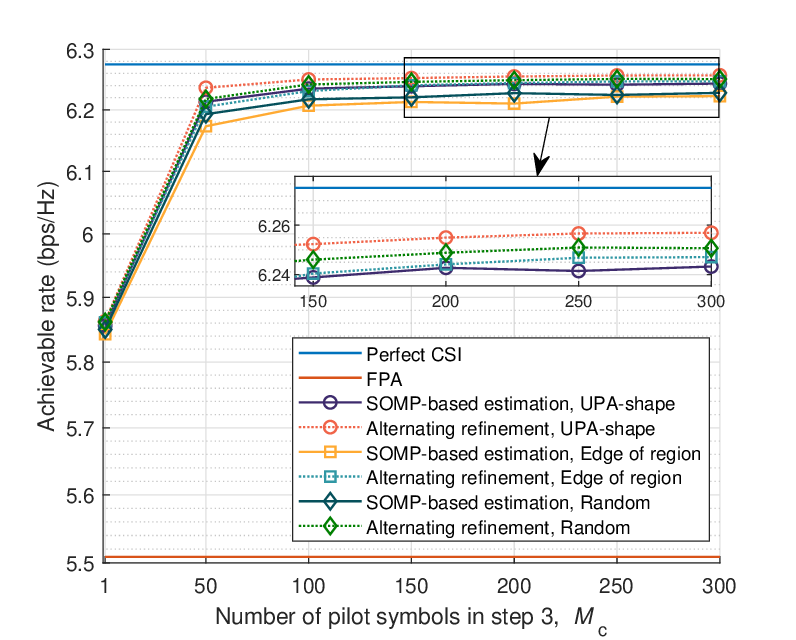}
		\caption{Achievable rates of the SOMP-based channel estimation method and alternating refinement method versus number $M_{\rm c}$ of pilots symbols in step 3.}
		\label{fig:m_rate}
	\end{figure}

	Fig. \ref{fig:m_rate} evaluates the achievable rates of different MA measurement position setups with varying number $M_{\rm c}$ of pilot symbols in step 3. There always exists a gap between the achievable rates with estimated CSI and that with the perfect CSI due to the estimation error. However, the gap is less than $1\%$ of the achievable rate with the perfect CSI, while both the SOMP-based channel estimation and alternating refinement methods can bring at least $12\%$ achievable rate gains over the FPA system with enough $M_{\rm c}$, which verifies the effectiveness of MA systems and the proposed channel estimation method. 
	
\begin{figure}[t]
		\centering
		\includegraphics[width= 7 cm]{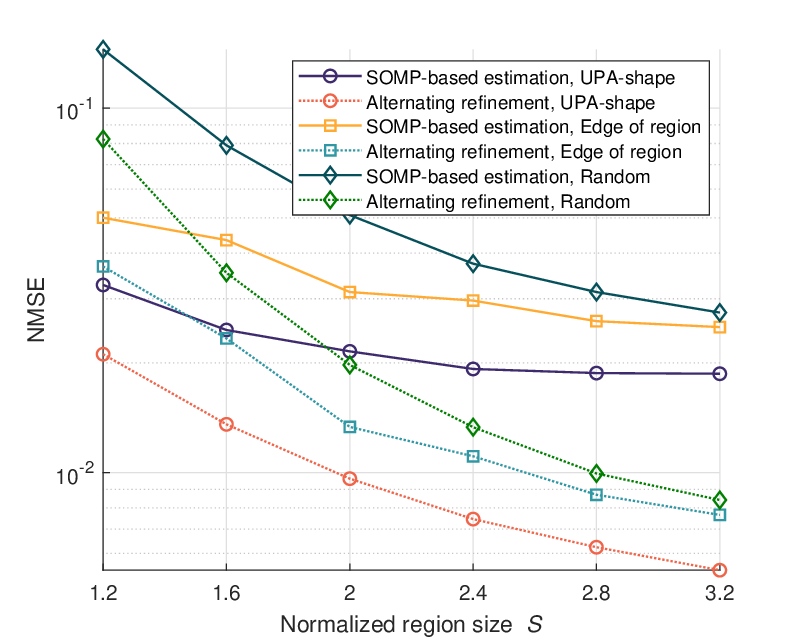}
		\caption{NMSE of the SOMP-based channel estimation method and alternating refinement method versus normalized region sizes $S$.}
		\label{fig:r_nmse}
	\end{figure}

	For MA systems, a large region leads to an increase of the hardware cost and power consumption due to the placement and movement of the MAs. In this regard, in Fig. \ref{fig:r_nmse}, we evaluate the channel estimation performance with varying region sizes. Notably, as the region becomes larger, more channel measurements are required to recover the angles, and thus we set $M_{\rm t} = M_{\rm r} = \left(S/0.4+1\right)^2$ while the total number for channel measurements remains $M_{\rm a} = 328$, which guarantees the distance between adjacent measurement positions is $0.4\lambda$ for UPA-shape scheme. As can be observed, the NMSE decreases with the increase of the region size. This is because a larger region size can reduce the correlation between different AoDs/AoAs, leading to a higher angular resolution \cite{10497534,10236898}. However, it is worth noting that a larger region size results in higher hardware cost and power consumption. On top of that, although the channel estimation performance with normalized region size $S = 1.2$ is worse in general, the NMSE can still be low with alternating refinement under the UPA-shape setup, which can outperform the SOMP-based channel estimation method with larger region sizes. Moreover, the UPA-shape setup yields the lowest NMSE. This demonstrates that the UPA-shape can also reduce the correlation of different AoDs/AoAs \cite{10236898}. 
	
\begin{figure}[t]
		\centering
		\includegraphics[width= 7 cm]{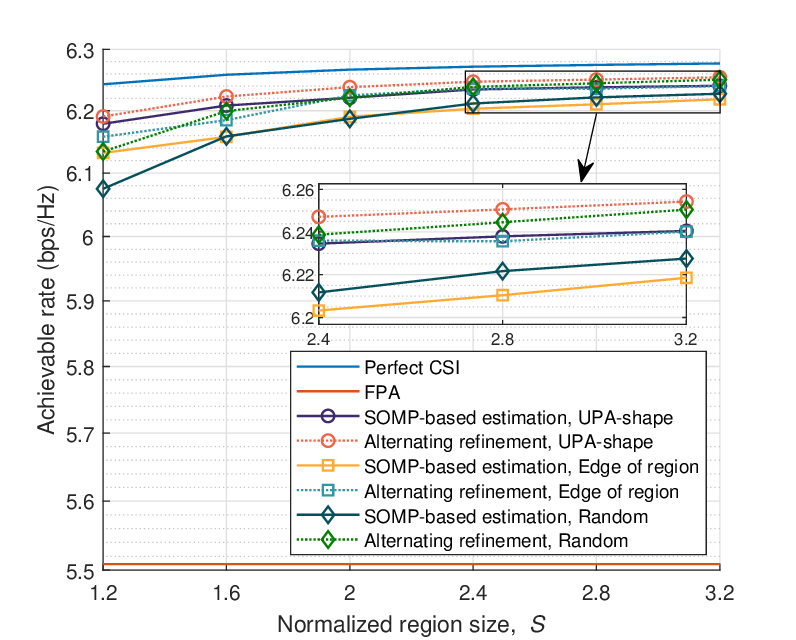}
		\caption{Achievable rates of the SOMP-based channel estimation method and alternating refinement method versus normalized region sizes $S$.}
		\label{fig:r_rate}
	\end{figure}

	Fig. \ref{fig:r_rate} shows the achievable rate versus varying normalized region size $S$ under different MA measurement position setups. The achievable rate with the perfect CSI increases slightly with the increase of normalized region size. This is because a larger achievable rate can be found in a larger region. Moreover, with normalized region size $S=1.2$, the achievable rates under the edge of region and random setups are lower due to the high estimation error, as shown in Fig. \ref{fig:r_nmse}. However, the UPA-shape setup can still obtain a comparable achievable rate to the perfect CSI, which yields the superior of the UPA-shape setup. In addition, although the NMSE of the edge of region setup is lower than the random setup, the edge of region setup obtains the lowest achievable rate among the three setups. This is because the edge of region setup performs a large number of channel measurements at the boundaries of the regions, which may lead to a large error of the CFRs in the remaining parts of the region. In this regard, the positions with the maximum achievable rate based on the edge of region setup may not be as accurate as the other setups. 
	
	\section{Conclusion}\label{section_6}
	
	In this paper, we proposed an SOMP-based channel estimation method and an alternating refinement channel estimation method for wideband MA systems, in which the complete wideband CSI, i.e., the CFRs between any position pair within the Tx region and the Rx region across all subcarriers, can be reconstructed with high accuracy. In particular, we first decomposed the CFRs into the FRVs, DRV, and PRT, which exhibit the sparse characteristic and can be recovered via an affordable number of pilots. Then, based on the proposed pilot scheme, the estimation of the FRVs as well as DRV was formulated as the problem of MMV-CS, which was solved via the SOMP algorithm. Next, the estimation of the PRT was carried out via the LS method. Moreover, an alternating refinement method was developed to further enhance the precision of the estimated FRVs, DRV, as well as PRT, in which the gradient descent algorithm was carried out to minimize the discrepancy between the received pilots and those constructed by the estimated CSI. Finally, comprehensive simulations were carried out to evaluate the channel estimation performance of the SOMP-based method and alternating refinement method. The results showed that both proposed methods can reconstruct the complete wideband CSI with high accuracy. Moreover, the alternating refinement method is able to achieve better channel estimation performance, but with a higher complexity.
	
	\begin{appendices}
		
		\section{The matricization of PRT ${\bf \Sigma}$}\label{sec:matrix}
		
		According to the definition of tensor multiplication, the CFR in \eqref{receive_pilot_single_c} can be rewritten as
			\begin{equation}\label{receive_pilot_single}
			\begin{small}
				\begin{aligned}
				& h\left({\bf t}_{\rm a}^{m_{\rm c}}, {\bf r}_{\rm a}^{m_{\rm c}}, k_{m_{\rm d}}\right)\\ = 
				& \left[\left({\bf \Sigma}\times_1{\bf f}\left({\bf r}_{\rm a}^{m_{\rm c}}\right)^{\rm H}\right)\times_2 {\bf g}\left({\bf t}_{\rm a}^{m_{\rm c}}\right)^{\rm T}\right]\times_3 {\bf d}\left(k_{m_{\rm d}}\right)^{\rm T}\\
				=&\sum_{l_{\rm d}=1}^{L_{\rm d}}\sum_{l_{\rm t} = 1}^{L_{\rm t}}\sum_{l_{\rm r} = 1}^{L_{\rm r}}\left[{\bf \Sigma}\right]_{l_{\rm r}, l_{\rm t}, l_{\rm d}}\times\left[{\bf g}\left({\bf t}_{\rm a}^{m_{\rm c}}\right)\right]_{l_{\rm t}}\times\left[{\bf f}\left({\bf r}_{\rm a}^{m_{\rm c}}\right)\right]_{l_{\rm r}}\times\left[{\bf d}\left(k_{m_{\rm d}}\right)\right]_{l_{\rm d}}\\
				=& \sum_{l_{\rm d}=1}^{L_{\rm d}}\left({\bf f}\left({\bf r}_{\rm a}^{m_{\rm c}}\right)^{\rm H}\left[{\bf \Sigma}\right]_{:,:,l_{\rm d}}{\bf g}\left({\bf t}_{\rm a}^{m_{\rm c}}\right)\right)\times\left[{\bf d}\left(k_{m_{\rm d}}\right)\right]_{l_{\rm d}}\\
				=& \sum_{l_{\rm d}=1}^{L_{\rm d}}\left[\left({\bf g}\left({\bf t}_{\rm a}^{m_{\rm c}}\right)^{\rm T}\otimes{\bf f}\left({\bf r}_{\rm a}^{m_{\rm c}}\right)^{\rm H}\right){\rm vec}\left(\left[{\bf \Sigma}\right]_{:,:,l_{\rm d}}\right)\right]\times\left[{\bf d}\left(k_{m_{\rm d}}\right)\right]_{l_{\rm d}}.
			\end{aligned}
			\end{small}
		\end{equation}
		Moreover, since ${\bf X} = \left[\left[\bf \Sigma\right]_{:,1,:}^{\rm T}, \cdots, \left[{\bf \Sigma}\right]_{:,L_{\rm t},:}^{\rm T}\right]^{\rm T}$, we have ${\rm vec}\left(\left[{\bf \Sigma}\right]_{:,:,l_{\rm d}}\right) = \left[{\bf X}\right]_{:,l_{\rm d}}$. Consequently, \eqref{receive_pilot_single} becomes
		\begin{equation}
			\begin{small}
				\begin{aligned}
				&h\left({\bf t}_{\rm a}^{m_{\rm c}}, {\bf r}_{\rm a}^{m_{\rm c}}, k_{m_{\rm d}}\right)\\
				=&\sum_{l_{\rm d}=1}^{L_{\rm d}}\left[\left({\bf g}\left({\bf t}_{\rm a}^{m_{\rm c}}\right)^{\rm T}\otimes{\bf f}\left({\bf r}_{\rm a}^{m_{\rm c}}\right)^{\rm H}\right)\left[{\bf X}\right]_{:,l_{\rm d}}\right]
				\times\left[{\bf d}\left(k_{m_{\rm d}}\right)\right]_{l_{\rm d}}\\
				=&\left({\bf g}\left({\bf t}_{\rm a}^{m_{\rm c}}\right)^{\rm T}\otimes{\bf f}\left({\bf r}_{\rm a}^{m_{\rm c}}\right)^{\rm H}\right){\bf X}{\bf d}\left(k_{m_{\rm d}}\right).
			\end{aligned}
			\end{small}
		\end{equation}
		Thus, step (a) in \eqref{receive_pilot_single_c} holds. 
	\end{appendices}
	
	\bibliographystyle{IEEEtran} 
	\bibliography{reference}

\begin{thebibliography}{10}
\providecommand{\url}[1]{#1}
\csname url@samestyle\endcsname
\providecommand{\newblock}{\relax}
\providecommand{\bibinfo}[2]{#2}
\providecommand{\BIBentrySTDinterwordspacing}{\spaceskip=0pt\relax}
\providecommand{\BIBentryALTinterwordstretchfactor}{4}
\providecommand{\BIBentryALTinterwordspacing}{\spaceskip=\fontdimen2\font plus
\BIBentryALTinterwordstretchfactor\fontdimen3\font minus
  \fontdimen4\font\relax}
\providecommand{\BIBforeignlanguage}[2]{{%
\expandafter\ifx\csname l@#1\endcsname\relax
\typeout{** WARNING: IEEEtran.bst: No hyphenation pattern has been}%
\typeout{** loaded for the language `#1'. Using the pattern for}%
\typeout{** the default language instead.}%
\else
\language=\csname l@#1\endcsname
\fi
#2}}
\providecommand{\BIBdecl}{\relax}
\BIBdecl

\bibitem{6736761}
E.~G. Larsson, O.~Edfors, F.~Tufvesson, and T.~L. Marzetta, ``Massive {MIMO}
  for next generation wireless systems,'' \emph{IEEE Commun. Mag.}, vol.~52,
  no.~2, pp. 186--195, February 2014.

\bibitem{8354789}
B.~Wang, F.~Gao, S.~Jin, H.~Lin, and G.~Y. Li, ``Spatial- and
  frequency-wideband effects in millimeter-wave massive {MIMO} systems,''
  \emph{IEEE Trans. Signal Process.}, vol.~66, no.~13, pp. 3393--3406, July
  2018.

\bibitem{4907446}
M.~Franceschetti and K.~Chakraborty, ``Space-time duality in multiple antenna
  channels,'' \emph{{IEEE} Trans. Wireless Commun.}, vol.~8, no.~4, pp.
  1733--1743, April 2009.

\bibitem{9704317}
D.~Pinchera and M.~D. Migliore, ``Low-cost antenna architectures with control
  of the local environment for {5G} and beyond {5G},'' in \emph{{IEEE} Int.
  Symp. Antennas Propagat. USNC-URSI Radio Sci. Meeting (APS/URSI)}, Dec 2021,
  pp. 797--798.

\bibitem{10045774}
B.~Ning, Z.~Tian, W.~Mei, Z.~Chen, C.~Han, S.~Li, J.~Yuan, and R.~Zhang,
  ``Beamforming technologies for ultra-massive mimo in terahertz
  communications,'' \emph{IEEE Open J. Commun. Soc.}, vol.~4, pp. 614--658,
  2023.

\bibitem{9724113}
A.~Pizzo, L.~Sanguinetti, and T.~L. Marzetta, ``Fourier plane-wave series
  expansion for holographic {MIMO} communications,'' \emph{{IEEE} Trans.
  Wireless Commun.}, vol.~21, no.~9, pp. 6890--6905, Sep. 2022.

\bibitem{10163760}
R.~Deng, Y.~Zhang, H.~Zhang, B.~Di, H.~Zhang, H.~V. Poor, and L.~Song,
  ``Reconfigurable holographic surfaces for ultra-massive {MIMO} in {6G}:
  Practical design, optimization and implementation,'' \emph{{IEEE} J. Sel.
  Areas Commun.}, vol.~41, no.~8, pp. 2367--2379, Aug 2023.

\bibitem{10417101}
J.~Zhu, Z.~Wan, L.~Dai, M.~Debbah, and H.~V. Poor, ``Electromagnetic
  information theory: Fundamentals, modeling, applications, and open
  problems,'' \emph{{IEEE} Wireless Commun.}, vol.~31, no.~3, pp. 156--162,
  June 2024.

\bibitem{zhu2024mimo}
J.~Zhu, V.~Y. Tan, and L.~Dai, ``{MIMO} capacity analysis and channel
  estimation for electromagnetic information theory,'' \emph{arXiv preprint
  arXiv:2406.04881}, 2024.

\bibitem{zhu2023can}
J.~Zhu, X.~Su, Z.~Wan, L.~Dai, and T.~J. Cui, ``Can electromagnetic information
  theory improve wireless systems? a channel estimation example,'' \emph{arXiv
  preprint arXiv:2310.12446}, 2023.

\bibitem{9500830}
R.~Jess~Williams, P.~Ramírez-Espinosa, E.~de~Carvalho, and T.~L. Marzetta,
  ``Multiuser {MIMO} with large intelligent surfaces: Communication model and
  transmit design,'' in \emph{{IEEE} Int. Conf. Commun.}, June 2021, pp. 1--6.

\bibitem{wong2020fluid}
K.-K. Wong, K.-F. Tong, Y.~Zhang, and Z.~Zhongbin, ``Fluid antenna system for
  {6G}: When bruce lee inspires wireless communications,'' \emph{Electronics
  Letters}, vol.~56, no.~24, pp. 1288--1290, 2020.

\bibitem{zhu2022modeling}
L.~Zhu, W.~Ma, and R.~Zhang, ``Modeling and performance analysis for movable
  antenna enabled wireless communications,'' \emph{{IEEE} Trans. Wireless
  Commun.}, vol.~23, no.~6, pp. 6234--6250, June 2024.

\bibitem{zhu2023mov}
L.~Zhu, W.~Ma, and R.~Zhang, ``Movable antennas for wireless communication: Opportunities and
  challenges,'' \emph{{IEEE} Commun. Mag.}, vol.~62, no.~6, pp. 114--120, June
  2024.

\bibitem{ma2022mimo}
W.~Ma, L.~Zhu, and R.~Zhang, ``{MIMO} capacity characterization for movable
  antenna systems,'' \emph{{IEEE} Trans. Wireless Commun.}, vol.~23, no.~4, pp.
  3392--3407, April 2024.

\bibitem{zhu2024historical}
L.~Zhu and K.-K. Wong, ``Historical review of fluid antenna and movable
  antenna,'' \emph{arXiv preprint arXiv:2401.02362}, 2024.

\bibitem{8060521}
S.~Basbug, ``Design and synthesis of antenna array with movable elements along
  semicircular paths,'' \emph{IEEE Antennas Wireless Propagat. Lett.}, vol.~16,
  pp. 3059--3062, 2017.

\bibitem{8041885}
\BIBentryALTinterwordspacing
\emph{{MEMS} Integrated and Micromachined Antenna Elements, Arrays, and Feeding
  Networks}.\hskip 1em plus 0.5em minus 0.4em\relax Wiley, 2008, pp. 829--865.
  [Online]. Available: \url{https://ieeexplore.ieee.org/document/8041885}
\BIBentrySTDinterwordspacing

\bibitem{ning2024movable}
B.~Ning, S.~Yang, Y.~Wu, P.~Wang, W.~Mei, C.~Yuen, and E.~Bjornson, ``Movable
  antenna-enhanced wireless communications: General architectures and
  implementation methods,'' \emph{arXiv preprint arXiv:2407.15448}, 2024.

\bibitem{10508218}
W.~Mei, X.~Wei, B.~Ning, Z.~Chen, and R.~Zhang, ``Movable-antenna position
  optimization: A graph-based approach,'' \emph{{IEEE} Wireless Commun. Lett.},
  vol.~13, no.~7, pp. 1853--1857, July 2024.

\bibitem{10414081}
G.~Hu, Q.~Wu, J.~Ouyang, K.~Xu, Y.~Cai, and N.~Al-Dhahir,
  ``Movable-antenna-array-enabled communications with {CoMP} reception,''
  \emph{{IEEE} Commun. Lett.}, vol.~28, no.~4, pp. 947--951, April 2024.

\bibitem{10458417}
S.~Yang, W.~Lyu, B.~Ning, Z.~Zhang, and C.~Yuen, ``Flexible precoding for
  multi-user movable antenna communications,'' \emph{IEEE Wireless Commun.
  Lett.}, vol.~13, no.~5, pp. 1404--1408, May 2024.

\bibitem{10447471}
Z.~Cheng, N.~Li, J.~Zhu, X.~She, C.~Ouyang, and P.~Chen, ``Enabling secure
  wireless communications via movable antennas,'' in \emph{IEEE Int. Conf.
  Acoust. Speech Signal Process.}, April 2024, pp. 9186--9190.

\bibitem{10416363}
G.~Hu, Q.~Wu, K.~Xu, J.~Si, and N.~Al-Dhahir, ``Secure wireless communication
  via movable-antenna array,'' \emph{IEEE Signal Process. Lett.}, vol.~31, pp.
  516--520, 2024.

\bibitem{10579873}
X.~Wei, W.~Mei, D.~Wang, B.~Ning, and Z.~Chen, ``Joint beamforming and antenna
  position optimization for movable antenna-assisted spectrum sharing,''
  \emph{{IEEE} Wireless Commun. Lett.}, 2024, {DOI}:
  {10.1109/LWC.2024.3421636}, 2024. ({E}arly access).

\bibitem{10437006}
X.~Chen, B.~Feng, Y.~Wu, D.~W. Kwan~Ng, and R.~Schober, ``Joint beamforming and
  antenna movement design for moveable antenna systems based on statistical
  {CSI},'' in \emph{IEEE Global Commun. Conf.}, Dec 2023, pp. 4387--4392.

\bibitem{xiao2023multiuser}
Z.~Xiao, X.~Pi, L.~Zhu, X.-G. Xia, and R.~Zhang, ``Multiuser communications
  with movable-antenna base station: Joint antenna positioning, receive
  combining, and power control,'' \emph{arXiv preprint arXiv:2308.09512}, 2023.

\bibitem{10354003}
L.~Zhu, W.~Ma, B.~Ning, and R.~Zhang, ``Movable-antenna enhanced multiuser
  communication via antenna position optimization,'' \emph{IEEE Trans. Wireless
  Commun.}, 2023, {DOI}: {10.1109/TWC.2023.3338626}, 2023. ({E}arly access).

\bibitem{cheng2023sum}
Z.~Cheng, N.~Li, J.~Zhu, X.~She, C.~Ouyang, and P.~Chen, ``Sum-rate
  maximization for movable antenna enabled multiuser communications,''
  \emph{arXiv preprint arXiv:2309.11135}, 2023.

\bibitem{sun2023sum}
Y.~Sun, H.~Xu, C.~Ouyang, and H.~Yang, ``Sum-rate optimization for ris-aided
  multiuser communications with movable antenna,'' \emph{arXiv preprint
  arXiv:2311.06501}, 2023.

\bibitem{10382559}
W.~Ma, L.~Zhu, and R.~Zhang, ``Multi-beam forming with movable-antenna array,''
  \emph{IEEE Commun. Lett.}, vol.~28, no.~3, pp. 697--701, March 2024.

\bibitem{10278220}
L.~Zhu, W.~Ma, and R.~Zhang, ``Movable-antenna array enhanced beamforming:
  Achieving full array gain with null steering,'' \emph{IEEE Commun. Lett.},
  vol.~27, no.~12, pp. 3340--3344, Dec 2023.

\bibitem{zhu2024dynamic}
L.~Zhu, X.~Pi, W.~Ma, Z.~Xiao, and R.~Zhang, ``Dynamic beam coverage for
  satellite communications aided by movable-antenna array,'' \emph{arXiv
  preprint arXiv:2404.15643}, 2024.

\bibitem{zhu2024performance}
L.~Zhu, W.~Ma, Z.~Xiao, and R.~Zhang, ``Performance analysis and optimization
  for movable antenna aided wideband communications,'' \emph{arXiv preprint
  arXiv:2401.08974}, 2024.

\bibitem{7454701}
Z.~Gao, C.~Hu, L.~Dai, and Z.~Wang, ``Channel estimation for millimeter-wave
  massive {MIMO} with hybrid precoding over frequency-selective fading
  channels,'' \emph{IEEE Commun. Lett.}, vol.~20, no.~6, pp. 1259--1262, June
  2016.

\bibitem{7174558}
Z.~Gao, L.~Dai, Z.~Wang, and S.~Chen, ``Spatially common sparsity based
  adaptive channel estimation and feedback for {FDD} massive {MIMO},''
  \emph{IEEE Trans. Signal Process.}, vol.~63, no.~23, pp. 6169--6183, Dec
  2015.

\bibitem{8846224}
A.~Liao, Z.~Gao, H.~Wang, S.~Chen, M.-S. Alouini, and H.~Yin, ``Closed-loop
  sparse channel estimation for wideband millimeter-wave full-dimensional
  {MIMO} systems,'' \emph{IEEE Trans. Commun.}, vol.~67, no.~12, pp.
  8329--8345, Dec 2019.

\bibitem{8306126}
J.~Rodríguez-Fernández, N.~González-Prelcic, K.~Venugopal, and R.~W. Heath,
  ``Frequency-domain compressive channel estimation for frequency-selective
  hybrid millimeter wave {MIMO} systems,'' \emph{IEEE Trans. Wireless Commun.},
  vol.~17, no.~5, pp. 2946--2960, May 2018.

\bibitem{7961152}
K.~Venugopal, A.~Alkhateeb, N.~González~Prelcic, and R.~W. Heath, ``Channel
  estimation for hybrid architecture-based wideband millimeter wave systems,''
  \emph{IEEE J. Sel. Areas Commun.}, vol.~35, no.~9, pp. 1996--2009, Sep. 2017.

\bibitem{zhang2023successive}
Z.~Zhang, J.~Zhu, L.~Dai, and R.~W. Heath~Jr, ``Successive {Bayesian}
  reconstructor for channel estimation in flexible antenna systems,''
  \emph{arXiv preprint arXiv:2312.06551}, 2023.

\bibitem{10236898}
W.~Ma, L.~Zhu, and R.~Zhang, ``Compressed sensing based channel estimation for
  movable antenna communications,'' \emph{{IEEE} Commun. Lett.}, vol.~27,
  no.~10, pp. 2747--2751, Oct 2023.

\bibitem{10497534}
Z.~Xiao, S.~Cao, L.~Zhu, Y.~Liu, B.~Ning, X.-G. Xia, and R.~Zhang, ``Channel
  estimation for movable antenna communication systems: A framework based on
  compressed sensing,'' \emph{IEEE Trans. Wireless Commun.}, {DOI}:
  {10.1109/TWC.2024.3385110}, 2024. ({E}arly access).

\bibitem{ruo2024channel}
Z.~Ruoyu, C.~Lei, Z.~Wei, G.~Xinrong, C.~Yueming, W.~Wen, and R.~Zhang,
  ``Channel estimation for movable-antenna {MIMO} systems via tensor
  decomposition,'' \emph{IEEE Wireless Commun. Lett.}, 2024, in {M}ajor
  revision.

\bibitem{3gpp38901}
3GPP TR 38.901: ``{S}tudy on channel model for frequencies from 0.5 to 100
  {GH}z", V16.1.0, 3rd Generation Partnership Project, Dec. 2019.

\bibitem{8765421}
L.~Zhang, L.~Huang, B.~Li, J.~Yin, and W.~Bao, ``Sensing matrix design for
  {MMV} compressive sensing: An {MVDR} approach,'' \emph{IEEE Trans. Veh.
  Tech.}, vol.~68, no.~9, pp. 8601--8612, Sep. 2019.

\end{thebibliography}
	
\end{document}